\documentclass[11pt]{article}
\usepackage{setspace}
\usepackage{graphicx}
\usepackage{lscape}
\usepackage{pdflscape}
\usepackage{afterpage}
\usepackage{pdfpages}
\usepackage{float}
\usepackage{booktabs}
\usepackage[utf8]{inputenc}
\usepackage{indentfirst}
\usepackage{arydshln}
\usepackage{arydshln}
\usepackage{tikz}
\usepackage{lipsum}
\usepackage{pgffor}
\usepackage{dsfont}
\usepackage{amsfonts}
\usepackage{amsmath}
\allowdisplaybreaks
\usepackage{mathtools}

\usepackage{xfrac}
\usepackage{amssymb}
\usepackage{bigints}
\usepackage{graphicx}
\usepackage{url}				
\usepackage{caption}
\usepackage{subcaption} 
\usepackage{enumitem}
\usepackage{relsize,exscale}
\usepackage{multirow}
\usepackage{natbib}
\setcitestyle{authoryear, open={(},close={)}}
\usepackage{datetime}

\usepackage{tabularx}
\newcolumntype{C}{>{\centering\arraybackslash}X}
\usepackage{pifont}

\newcommand{\xmark}{\ding{55}}

\newdateformat{monthyeardate}{%
	\monthname[\THEMONTH] \THEYEAR}

\usepackage{titlesec}
\titleformat*{\section}{\large\bfseries}
\titleformat*{\subsection}{\large\bfseries}


\newcounter{parentnumber}
\usepackage{theorem}

\providecommand{\U}[1]{\protect\rule{.1in}{.1in}}

\usepackage{hyperref} 
\hypersetup{
	colorlinks=true, breaklinks=true, bookmarks=true,bookmarksnumbered,
	urlcolor=blue, linkcolor=blue, citecolor=blue, 
	pdftitle={}, 
	pdfauthor={\textcopyright}, 
	pdfsubject={}, 
	pdfkeywords={}, 
	pdfcreator={pdfLaTeX}, 
	pdfproducer={LaTeX with hyperref and ClassicThesis} 
}

\begin{document}
	\setstretch{1}
\title{{\LARGE DETERring more than Deforestation: Environmental Enforcement Reduces Violence in the Amazon\thanks{We would like to thank Lucas Belleza, Lucas Finamor, Livia Haddad, Rafael Pucci, Leila Pereira, Rodrigo Soares, and Robert Heilmayr for helpful suggestions. We also thank seminar participants at Sao Paulo School of Economics - FGV, 2025 RIDGE Forum on Environmental Economics, and FEA-USP Workshop on the Economics of Crime. The paper uses pre-existing public observational data, so pre-registration and a pre-analysis plan were not feasible; we therefore note that the study is exempt from these requirements.}}}

\author{
	Rafael Araujo\thanks{Sao Paulo School of Economics - FGV. Email: \href{mailto:rafael.araujo@fgv.br}{rafael.araujo@fgv.br}. The author thanks Rede de Pesquisa em Produtividade e Sustentabilidade (Rede PPS) for the financial support. The funders had no role in the study design, data collection and analysis, interpretation of results, manuscript preparation, or the decision to submit the article for publication.}  \and Vitor Possebom\thanks{Sao Paulo School of Economics - FGV. Email: \href{mailto:vitor.possebom@fgv.br}{vitor.possebom@fgv.br}. This study was financed, in part, by the São Paulo Research Foundation (FAPESP), Brazil. Process Number \#2025/04857-0.}   \and Gabriela Setti\thanks{Sao Paulo School of Economics - FGV. Email: \href{mailto:gabriela.coutinho@fgv.edu.br}{gabriela.coutinho@fgv.edu.br}}
}
\date{}

\maketitle

\newsavebox{\tablebox} \newlength{\tableboxwidth}


\begin{center}



	%
	%

	\

	\large{\textbf{Abstract}}
\end{center}

We estimate the impact of environmental law enforcement on violence in the Brazilian Amazon. The introduction of the Real-Time Deforestation Detection System (DETER), which enabled the government to monitor deforestation in real time and issue fines for illegal clearing, significantly reduced homicides in the region. To identify causal effects, we exploit exogenous variation in satellite monitoring generated by cloud cover as an instrument for enforcement intensity. Our estimates imply that the expansion of state presence through DETER prevented approximately 1,477 homicides per year,  a 15\% reduction in homicides. These results show that a replicable environmental enforcement policy produces social benefits.
\


\


\newpage

\doublespacing

\section{Introduction}\label{SecIntro}

In regions marked by weak state presence, insecure property rights and widespread illegal markets, disputes over land, resources and markets often escalate into violence \citep{collier2004greed,angrist2008rural,dell2015trafficking}. Strengthening state capacity and law enforcement can revert this scenario \citep{becker1968crime,besley2010state}. In the Brazilian Amazon --- where state presence has recurrently been exercised through environmental law enforcement --- policies designed to curb deforestation may also reduce violence, generating security co-benefits and challenging a presumed trade-off between development and environmental protection \citep{foster2003economic,stern2017environmental,jayachandran2022economic}.

This paper investigates whether environmental law enforcement can reduce homicides in the Brazilian Amazon, a region where illegal resource extraction drives both deforestation and persistent conflict \citep{chimeli2017use,fetzer2017take,pereira2022tale}. We focus on the introduction of the Real-Time Deforestation Detection System (DETER), a satellite-based monitoring and enforcement policy implemented to curb deforestation. While previous studies have documented that DETER significantly reduces deforestation \citep{assuncao2023deter}, its broader societal effects remain largely unexplored.

We find that the intensified enforcement and state presence associated with DETER reduced violence in the Amazon. This evidence is important because it shows that strengthening environmental monitoring capacity need not generate adverse development outcomes. The deployment of DETER---arguably the most consequential innovation in Brazilian environmental policy---decreased the homicide rate in the Brazilian Amazon. This finding has broader implications for strengthening enforcement in environmentally sensitive regions across the developing world.

This empirical result is of interest because the effect of environmental enforcement on violence is theoretically ambiguous. On the one hand, by increasing the cost of deforestation through a higher probability of fines and apprehensions, enforcement can reduce incentives to engage in violent competition over land and resource extraction. Moreover, if deforestation is an input for other illegal activities, such as mining or constructing airstrips for drug trafficking, enforcement raises the costs of these activities as well. On the other hand, suppressing profitable activities may generate negative income shocks for individuals or groups reliant on them, potentially pushing them toward alternative illicit activities. In such cases, enforcement could inadvertently increase certain forms of violence \citep{macleod2023crime,macleod2024positive}.

Understanding which mechanism dominates is crucial for policy design. If environmental enforcement reduces violence, it strengthens the case for integrated strategies that promote both conservation and public security. Conversely, if it exacerbates violence, complementary interventions may be necessary to offset unintended consequences \citep[e.g.,][]{soares2004development,tuttle2019snapping,deshpande2022does}. Our results, by indicating that the net effect of environmental protection is crime-reducing, support the case for integrated strategies.

To identify causal effects, we exploit a key technical constraint of satellite-based enforcement: deforestation can only be sanctioned when it is visible from space. Cloud coverage introduces quasi-random interruptions in this monitoring technology, weakening enforcement. We use this variation as an instrument for enforcement, as in prior work leveraging satellite visibility constraints to study deforestation \citep{assuncao2023deter}. We explore this relationship during the stable implementation period of DETER for environmental enforcement (2006–2016) --- after which the system was gradually dismantled, particularly under President Bolsonaro's administration (2019-2022). In the first stage, cloud coverage predicts the intensity of enforcement actions, measured by the number of fines issued. In the second stage, the predicted enforcement intensity is used to estimate its effect on municipal homicide rates.

The results show that intensified environmental enforcement significantly reduces violence in the Brazilian Amazon. One additional deforestation-related fine is associated with a decrease of approximately 0.73 homicides per 100,000 inhabitants --- a reduction equivalent to 2.58\% relative to the sample mean homicide rate of 28.16. An increase in enforcement intensity from the 25th to the 75th percentile of the fines distribution (approximately 8 additional fines per year) corresponds to a substantial reduction of 5.82 homicides per 100,000 inhabitants, representing a 20.7\% decline from the average observed violence level.

When we scale these estimated benefits by the average enforcement intensity and the Amazon's population size, we find that DETER prevented 1,477 homicides per year. This decrease represents a 15\% reduction from the annual average of 8,790 homicides in the region. Combining this reduction with policy costs and willingness-to-pay estimates from the literature, we find that, even ignoring its environmental objectives, DETER’s law enforcement component alone yields a benefit–cost ratio of at least 3.7. These results underscore the substantial deterrent effect of environmental enforcement in regions characterized by weak institutions and contested land use.

To strengthen the causal interpretation of our findings, we conduct several robustness and sensitivity analyses. In particular, distributional regressions indicate that environmental enforcement significantly reduces not only the intensity but also the incidence and endemic nature of lethal violence across municipalities. Specifically, each additional environmental fine reduces the probability that a municipality records any homicide by 1.2 percentage points. This result is stronger in municipalities where violence is endemic: each additional fine lowers the probability of recording a homicide rate above 10 by 1.8 percentage points. Our results are robust to potential underreporting of homicides and to alternative measures of violence. We further validate our instrumental variable approach by applying the sensitivity analysis method proposed by \citet{cinelli2025omitted}, finding that our main conclusions are not overturned by cofounding factors of similar magnitude to key observable covariates.

We also show that the reduction in violence is concentrated among black adult men, a group that bears a systematically disproportionate burden of lethal violence in Brazil. This pattern is consistent with longstanding racial disparities in homicide victimization documented in the country: black Brazilians account for the overwhelming majority of homicide victims \citep{bfps2021}.

This study contributes to three strands of the economics literature. Concerning environmental law enforcement, we highlight a co-benefit: in settings with weak institutions and poorly defined property rights, environmental protection and development need not be opposing goals. Previous work has demonstrated the effectiveness of environmental policies in reducing deforestation in the Amazon (\citealp{assunccao2020effect}; \citealp{assuncao2023deter}; \citealp{assuncao2023optimal}; \citealp{bragancca2022cutting}). However, their broader societal implications remained largely unexplored. By analyzing the impact of environmental enforcement on homicide rates, we document a substantial co-benefit of these policies, showing that efforts to curb illegal deforestation also generate meaningful reductions in local violence. This perspective underscores that environmental enforcement yields multidimensional benefits, which should be incorporated into policy evaluation and cost–benefit analysis.

Thus, while one key rationale for curbing deforestation is to address a global externality (carbon emissions), our results indicate that accounting for reductions in violence not only increases the estimated social benefits of environmental enforcement but also shifts the incidence of those benefits from global to local. To the extent that conservation policies generate tangible local benefits, they may also be more politically sustainable, as they mitigate the concern that Brazil bears the economic costs of conservation (e.g., foregone agricultural income) while the benefits are diluted across the global community through carbon storage.

Second, our paper extends a large literature on crime determinants and deterrence policies. Foundational work has emphasized crime as a rational choice problem shaped by deterrence incentives \citep{becker1968crime,ehrlich1973participation}, while other studies have provided evidence on the effectiveness of policing and enforcement in reducing crime \citep{levitt2002using,di2004police,draca2011panic}. Of particular interest is the literature on violence in contestable markets \citep{angrist2008rural,dube2013commodity,idrobo2014illegal}. Specifically in the literature on violence in the Brazilian Amazon, \cite{fetzer2017take} shows that the homologation of indigenous territories reduces violence, while other papers focus on the effects of particular illegal markets, such as the mahogany trade \citep{chimeli2017use}, illegal gold mining \citep{pereira2022tale}, or drug trafficking \citep{pereira2024landing}. In contrast, our analysis evaluates a general anti-deforestation enforcement policy implemented throughout the entire Amazon biome without the explicit objective of reducing violence.

Third, this paper contributes to the broader literature on deforestation. This literature has examined determinants of deforestation and evaluated policies to curb it, including carbon taxes \citep{souza2019deforestation,araujo2025efficient}, trade restrictions \citep{dominguez2021efficiency,hsiao2021coordination,farrokhi2025deforestation}, and credit policies \citep{assunccao2020effect}. Our study shifts the focus from environmental outcomes to social ones. This perspective broadens the scope of conservation benefits beyond carbon emissions, underscoring that in a weak institutional environment the social value of conservation is larger than previously recognized.\footnote{A complementary literature examines development-oriented policies and finds that, while cash transfer programs may increase deforestation \citep{rocha2023relationship}, they may reduce deforestation when made conditional on conservation \citep{wong2022individual,cisneros2022impacts}.}

The remainder of the study is structured as follows. Section~\ref{SecBack} presents a detailed institutional background on deforestation, violence, and environmental enforcement in the Amazon. Section~\ref{SecModel} presents a conceptual model to discuss potential effects of environmental enforcement on violence. Section~\ref{SecData} describes data sources and descriptive statistics. Section~\ref{SecEmpirical} outlines the empirical approach, describing the identification strategy and econometric methods. Section~\ref{SecResults} discusses the main empirical findings and discuss potential mechanisms. Section~\ref{SecRobChecks} presents a series of robustness checks, extensions, and sensitivity analyses that assess the validity and stability of the main results. Section~\ref{SecConclusion} concludes with a summary of key findings and policy implications.

\section{Crime, Environment, and Governance}\label{SecBack}
This section provides background on the three core elements of the analysis: violence, deforestation, and environmental policies.\footnote{
	A detailed discussion of each of these elements is presented in Appendix~\ref{AppBack}.}

\subsection{Violence in the Amazon}

Crime can emerge from strategic responses to institutional weaknesses. When the rule of law is fragile and enforcement capacity limited, illegal markets may rely on coercion to enforce contracts, resolve disputes, and claim territorial control. These dynamics are especially pronounced in the Brazilian Amazon, characterized by land tenure insecurity and weak state presence. Unlike most of Brazil, where violence tends to concentrate in urban areas, the Amazon exhibits a rural pattern of lethal violence --- particularly near deforestation frontiers, contested landholdings, and zones of illegal mining and logging \citep{hrw2019}. These patterns point to a distinctive configuration of violence shaped by the interplay between environmental degradation and institutional fragility \citep{chimeli2017use, de2024environmental}.

\begin{figure}[ht]
	\centering
	\caption{Homicide Rate: Municipalities in the Amazon Biome vs. Outside the Biome}
	\label{fig:homicide_trend}
	\includegraphics[width=0.75\linewidth]{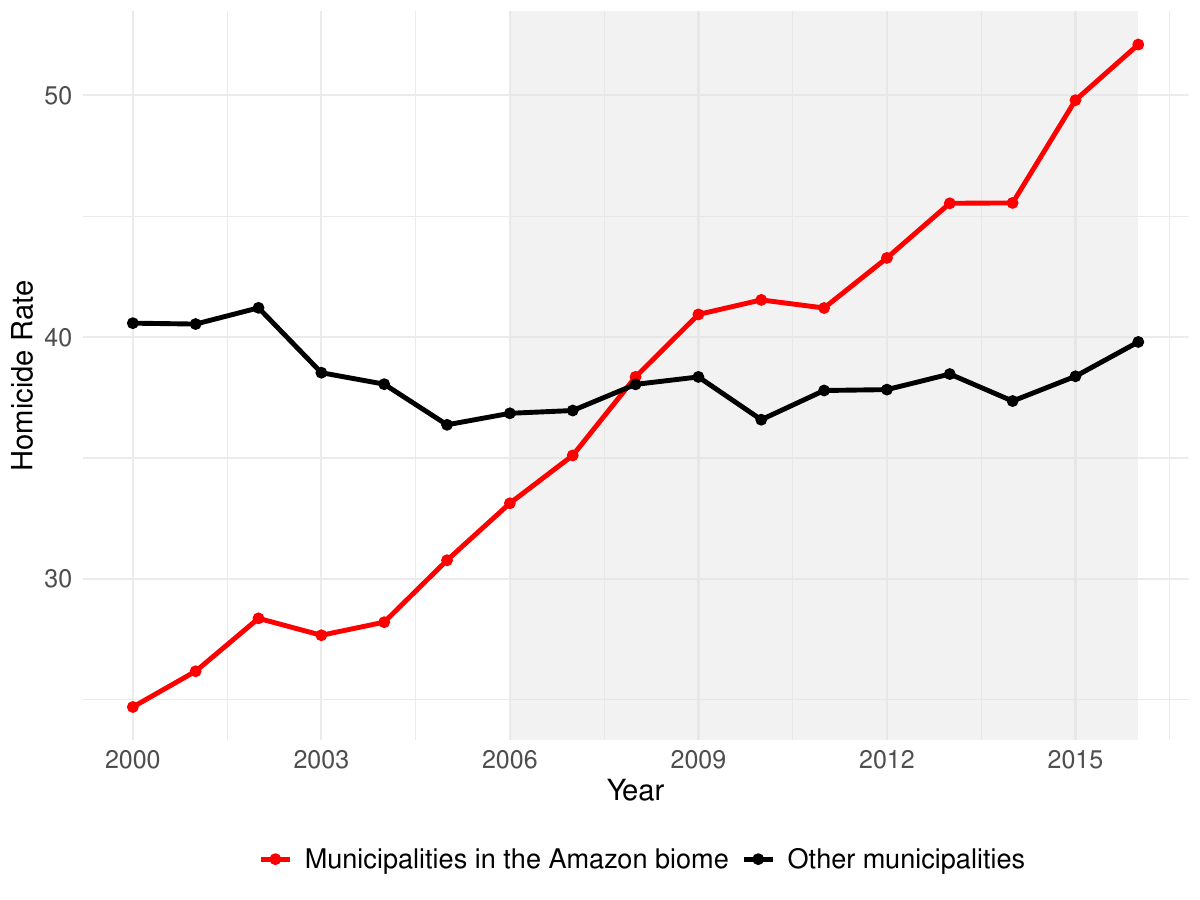}
\end{figure}

In recent decades, homicide trends in the Amazon diverged sharply from the rest of the country. While national rates of lethal violence declined, Amazon municipalities experienced a sustained increase. Figure~\ref{fig:homicide_trend} shows that by the late 2000s, homicide rates in the Amazon had surpassed those of non-Amazon areas --- a gap that continued to widen over time. Between 2006 and 2016, the average homicide rate in Amazon municipalities rose from 33.1 to 52.1 per 100,000 inhabitants, a 57.3\% increase. In contrast, non-Amazon municipalities saw only a 8.0\% rise. This divergence highlights how environmental crime and institutional fragility have created a persistent and regionally concentrated pattern of violence in the Brazilian Amazon \citep{fbsp2021amazon, ipea2025}.\\

\subsection{Deforestation: Dynamics and Drivers}

Deforestation in the Brazilian Amazon has reached alarming levels in recent decades, with more than 15\% of the original forest cover already cleared \citep{inpeTerrabrasilis}. Most of this clearing has been illegal \citep[see][]{mapbiomas2023rad}, often occurring in areas where all deforestation is prohibited --- such as protected lands, indigenous territories, and undesignated public forests \citep{assuncao2023deter, moutinho2023untitled}. Even in zones where deforestation could be legal in principle, noncompliance with environmental regulations is widespread \citep{godar2012responsible,rajao2020rotten}.

These patterns are deeply rooted in institutional fragility, unresolved land tenure, and public policies that historically encouraged frontier expansion without addressing overlapping claims \citep{fearnside2005deforestation, barreto2008dono}. Land grabbing for agricultural use remain the main driver of deforestation. In regions with weak enforcement and legal uncertainty, violence is a mean of asserting control and pressuring the state to regularize illegal occupations \citep{alston1999model, alston2000land, angrist2008rural, idrobo2014illegal, sant2021environmental}.\\

\subsection{Environmental Law Enforcement and Monitoring}

In response to growing environmental degradation and international pressure, the Brazilian government progressively strengthened its deforestation control policies throughout the 2000s. A turning point came in 2004 with the launch of the Action Plan for the Prevention and Control of Deforestation in the Legal Amazon (Plano de Ação para Prevenção e Controle do Desmatamento na Amazônia Legal, or PPCDAm), which combined land-use regulation, environmental enforcement, and technological innovation to curb deforestation in the region. The plan marked a shift away from reactive and fragmented responses toward an integrated and proactive strategy, with a particular emphasis on improving monitoring capabilities and enhancing the capacity of enforcement agencies to detect and respond to illegal forest clearing.

A central innovation of the PPCDAm was the creation of the Real-Time Detection of Deforestation System (DETER), a satellite-based system developed by the Brazilian Institute for Space Research (INPE) to provide real-time alerts of deforestation activity. By processing high-frequency satellite imagery, DETER identifies recent forest disturbances and produces georeferenced alerts that guide inspection teams in the field \citep{inpe_deter_modis_2004_2017}. Because enforcement is more effective when infractions are caught in the act, timely alerts significantly increase the likelihood of sanctioning violators --- raising the perceived probability of detection and thereby strengthening deterrence \citep{ferreira2023satellites}. Although cloud cover and limited local capacity still pose challenges, the system has proven effective in reducing deforestation \citep{assuncao2023deter,vieira2023role,picchetti2024instrumental}.

\section{Conceptual Framework}\label{SecModel}

In this section, we discuss why environment enforcement may have an ambiguous effect on violence. We build an occupational choice model in the spirit of Roy, where individuals choose among three activities: work not related with deforestation, deforestation, and crime unrelated to the deforestation activity (which we denote simply as crime). The payoffs are given by $\pi^j=r_j \varepsilon^j$ for $j \in \{\text{work},\text{def},\text{crime}\}$ in the form of:
\[
\pi^{work} = w_{work}\varepsilon^{work}, \qquad
\pi^{def} = (w_{def} - \theta c_{def})\varepsilon^{def}, \qquad
\pi^{crime} = (w_{crime} - \gamma c_{crime})\varepsilon^{crime},
\]
where $w_{j}$ is the baseline return to activity $j \in \{\text{work},\text{def},\text{crime}\}$; $c_{def}$ and $c_{crime}$ are punishments if caught; $\theta,\gamma \in [0,1]$ are the probabilities of detection for deforestation and crime; $\varepsilon^{j}$ are idiosyncratic shocks; let $r_{work} \coloneqq w_{work}$, $r_{def} \coloneqq (w_{def} - \theta c_{def})$ and $r_{crime} \coloneqq (w_{crime} - \gamma c_{crime})$.

The agent chooses the occupation with the highest payoff:
\[
\max \left\{ \pi^{work}, \pi^{def}, \pi^{crime} \right\}.
\]

\textbf{Violence. } Suppose deforestation generates violent conflict with probability $P_{def}$, and crime with probability $P_{crime}$. Under the standard assumption that the idiosyncratic shocks follow a Frechet distribution, the share of each occupation is given by

\[
s_{j} = \frac{r_j}{r_{work} + r_{def} + r_{crime}} \quad \text{for } j \in \{\text{work},\text{def},\text{crime}\}
\]

We can then compute expected violence as
\[
V = s_{def} P_{def} + s_{crime} P_{crime}
\]

\textbf{Comparative Statics. } The effect of enforcement against deforestation is captured by the derivative of $V$ with respect to~$\theta$:
\[
\frac{\partial V}{\partial \theta}
= \frac{\partial s_{def}}{\partial \theta} P_{def}
+ \frac{\partial s_{crime}}{\partial \theta} P_{crime},
\]
implying that
\[
\frac{\partial V}{\partial \theta}
\propto c_{def}\left[r_{crime}P_{crime} -(r_{work}+r_{crime})P_{def}\right]
\]

Consequently, increasing enforcement can either raise or reduce violence. In particular, violence decreases if and only if
\[
\frac{r_{crime}}{r_{work}+r_{crime}} < \frac{P_{def}}{P_{crime}}.
\]
Intuitively, violence decreases when the relative return to crime is small compared with the relative propensity of deforestation to generate violence, and it increases when the inequality is reversed. A setting with weak institutions is captured by a higher value of $P_{def}$. When $P_{def}=0$, disputes over deforestation are resolved through formal institutions rather than violence. In this case, punishing deforestation simply reduces the payoff of an otherwise legal activity, thereby shifting individuals toward crime through an income effect.

\section{Data}\label{SecData}
The empirical analysis relies on a municipality-year panel from 2006 to 2016, covering municipalities located entirely or partially within the Amazon biome. We adopt the same sample of \cite{assuncao2023deter}, using the 2007 administrative boundaries defined by the Brazilian Institute for Geography and Statistics (IBGE) as a consistent reference throughout the sample period. The dataset comprises 521 municipalities.


Satellite deforestation monitoring system's definition of “year” differs from the calendar year. Specifically, in the deforestation measure from INPE’s Project for Monitoring Deforestation in the Legal Amazon (PRODES), year $t$ spans from August of year $(t-1)$ to July of year $t$. As such, detection alerts from DETER are also aggregated following this definition of the PRODES year. For sources reporting monthly data, we aggregate observations to align with the PRODES year. For sources reporting data annually, we use values from calendar year $(t-1)$ to match PRODES year $t$. Throughout this analysis, all references to years correspond to PRODES years, unless otherwise noted.

A more technical discussion of variable construction and additional documentation can be found in Appendix~\ref{AppData}.

\subsection{Homicide Rate}
\label{sec:data_homicide_rate}
Measuring crime is inherently challenging due to data limitations, particularly in remote and underserved regions such as the Brazilian Amazon.\footnote{
	For discussions on the difficulties of crime measurement in the Amazon, see \citet{fbsp2021amazon}.
} In this context, the homicide rate per 100,000 inhabitants has become a widely accepted measure for violence and criminal activity in empirical research \citep{chimeli2017use, pereira2022tale, de2024environmental}. Homicides are more likely to be reported than other crimes and are generally recorded consistently across time and space. The homicide rate of municipality $i$ in year $t$ is calculated as the number of homicides per 100,000 inhabitants.

In Brazil, official mortality records are maintained by the Mortality Information System (SIM), part of the national Unified Health System (SUS). SIM data are made publicly available through the DataSUS platform, which provides detailed health statistics, including deaths and causes of death. SIM collects information based on death certificates, which are completed by health professionals and include both natural and external causes of death.

Homicide data in the Amazon face some challenges. Poor classification of causes of death --- especially under “ill-defined” or “undetermined intent” categories --- can compromise data quality and contribute to the underreporting of violent deaths \citep{ipea2024homicidios}. To address this challenge, \cite{ipea2025} uses the concept of hidden homicides by combining an inclusive coding strategy with a machine learning algorithm. Since data on hidden homicides does not exist at the municipal level for our sample period, we circumvent this issue by adopting an even more inclusive coding strategy and considering three categories for cause-of-death (ICD-10): assaults by various means (X85--Y09), events of undetermined intent (Y10--Y34), and intentional self-harm (X60--X84). We also show results for alternative categories in Section \ref{SecRobChecks}. To compute the homicide rate, we also need municipal population data. We use annual population estimates provided by Brazilian Institute of Geography and Statistics (IBGE). For municipality-years with missing population data, we apply spline interpolation.

\subsection{Law Enforcement}
\label{sec:data_enforcement}
Measuring environmental law enforcement is difficult, largely due to the lack of consistent and disaggregated administrative data at the municipal level. As in \cite{assuncao2023deter}, we use the total number of deforestation-related fines issued by the Brazilian Institute of Environment and Renewable Natural Resources (IBAMA) in each municipality and year as a proxy for the presence of enforcement. IBAMA provides publicly available electronic database containing detailed records on all environmental fines issued in Brazil.

\subsection{Cloud Coverage}
\label{sec:data_cloud_coverage}
A key limitation of the DETER system is its reliance on satellite imagery, which can be severely obstructed by cloud cover and other atmospheric conditions. Although environmental monitoring occurs at high frequency, enforcement is only possible when infractions are visible from space. Since cloud cover directly affects the system’s ability to detect illegal activity, the DETER dataset systematically reports monthly cloud coverage, including precise information on the spatial extent of observable areas.

Following \citet{assuncao2023deter}, we use the annual municipality-level measure of cloud coverage constructed from DETER data, obtained from the authors’ replication package \citep{assuncao2023code}. This variable reflects the average monthly ratio of cloud-covered area to total municipal area, capturing variation in satellite visibility due to weather conditions.

\subsection{Other Controls}
\label{sec:data_controls}
The set of control variables is organized into three categories: (i) satellite-based controls; (ii) weather-related controls; and (iii) socioeconomic controls. This subsection presents the data sources, definitions, and construction procedures for each category.

\subsubsection{Satellite Controls}
\label{sec:data_satellite_controls}
To account for potential measurement limitations in satellite-based monitoring, we include two municipality-level indicators of obstructions to PRODES imagery, which is the system that provides the official annual measure of deforestation, provided by INPE. The first measures the proportion of municipal area covered by clouds during each monitoring period. The second captures the share classified as non-observable due to atmospheric interference, such as cloud shadows or smoke from forest fires. Although the identification strategy relies on variation in DETER visibility, controlling for PRODES obstructions helps address concerns that persistent limitations in satellite-based monitoring --- across both systems --- may reflect broader structural constraints in state capacity or environmental governance. These constraints could influence the allocation of enforcement resources or correlate with violence-related dynamics. Including these controls helps ensure that the variation in DETER cloud coverage exploited for identification is not confounded by latent monitoring gaps that shape institutional responses or local conflict environments.

\subsubsection{Weather Controls}
\label{sec:data_weather_controls}
Climatic conditions are relevant to both environmental enforcement, local economic and social dynamics in the Amazon region. Following \citet{assuncao2023deter}, we include two annual weather indicators: total precipitation and average temperature. These variables are based on the monthly gridded datasets compiled by \citet{matsuura2018precipitation} and \citet{matsuura2018temperature}, with a spatial resolution of 0.5 degrees by 0.5 degrees. For each municipality and year, total precipitation is computed as the sum of monthly rainfall, while average temperature is the mean of monthly air temperatures.

\subsubsection{Socioeconomic Controls}
\label{sec:data_socioeconomic_controls}
We include four main controls based on municipal-level data: (i) commodity price index, (ii) gross domestic product, (iii) population density, and (iv) education quality.

\textbf{Commodity price index. } Following \citet{assunccao2015deforestation}, we construct an output price series designed to capture exogenous variation in demand for agricultural commodities relevant to the region. Building on \citet{assuncao2023deter}, we construct a weighted real price index for six major commodities --- beef cattle, soybeans, cassava, rice, corn, and sugarcane.\footnote{
	Together, these land uses account for nearly 85\% of the agricultural land in sample municipalities during the study period.
} For beef cattle, weights reflect the 2004–2005 average ratio of cattle heads to municipal area in each municipality, using data from the Municipal Livestock Survey (PPM/IBGE). For crops, weights are based on the 2004–2005 average ratio of farmland to municipal area for each commodity $c$ in municipality $i$, using data from the Municipal Crop Production Survey (PAM/IBGE).

\textbf{Gross Domestic Product. } Municipal gross domestic product (GDP) is obtained from IBGE and reflects the total economic output of each municipality in a given year.

\textbf{Population density. } Population density is calculated as the ratio between total population and the area of each municipality (measured in square kilometers), using data from IBGE.

\textbf{Education quality. } Educational outcomes are proxied by the Basic Education Development Index (IDEB), published biennially by the National Institute for Educational Studies and Research Anísio Teixeira (INEP). The IDEB is one of the most widely used indicators of public school performance in Brazil. This index combines standardized test scores (Prova Brasil/Saeb) and school promotion rates, producing a composite score between 0 and 10 --- with higher values indicating better performance.

We use the IDEB score for early primary education (grades 1–5), which has broad coverage and comparability across municipalities and years. This choice reflects the focus on Brazil’s constitutionally mandated compulsory education and ensures a consistent proxy for basic human capital across locations. Since IDEB data are available only for even-numbered years and some municipalities have missing values, we construct an annual series by applying spline interpolation.

\subsection{Descriptive Statistics}
\label{sec:data_descriptive_statistics}
\begin{figure}[!htb]
	\centering

	\caption{Geographical distribution of key variables in 2006 within the Amazon biome}

	\label{fig:maps_2006}

	\begin{subfigure}[t]{0.48\textwidth}
		\includegraphics[width=\textwidth]{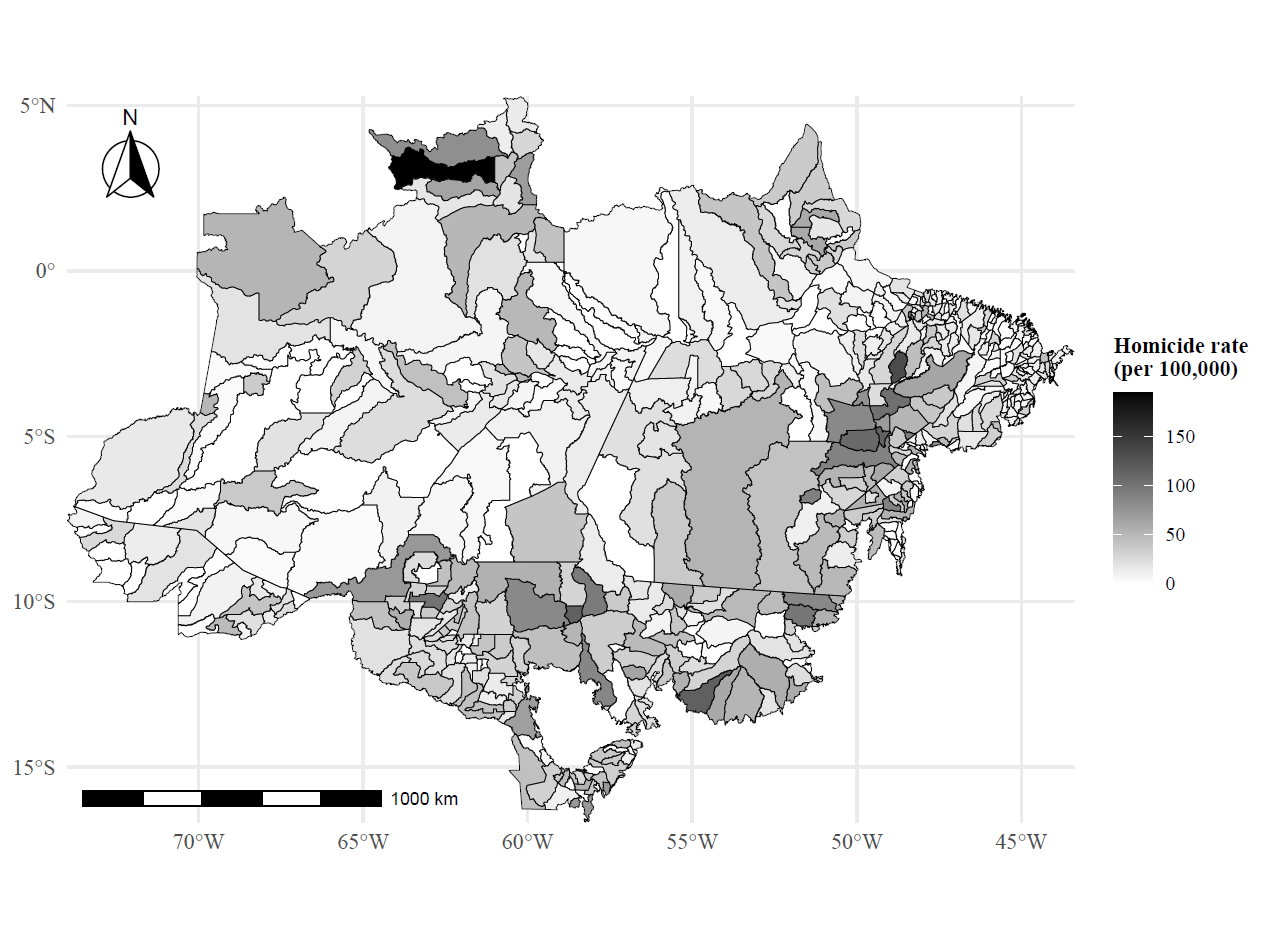}
		\caption{Homicide Rate}
		\label{fig:homicide_2006}
	\end{subfigure}
	\hfill
	\begin{subfigure}[t]{0.48\textwidth}
		\includegraphics[width=\textwidth]{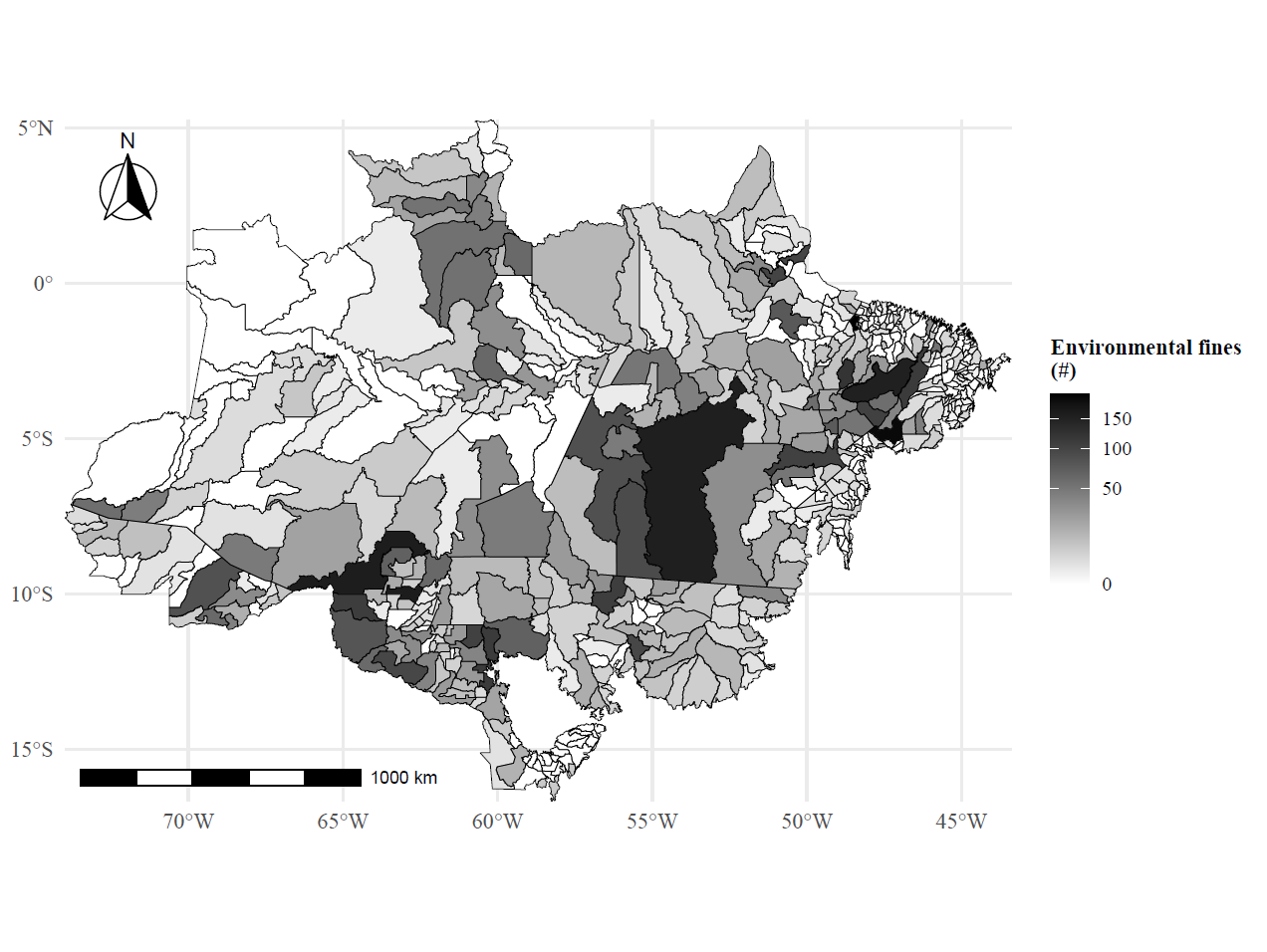}
		\caption{Environmental Enforcement}
		\label{fig:enforcement_2006}
	\end{subfigure}

	\vspace{0.5em}

	\begin{subfigure}[t]{0.48\textwidth}
		\centering
		\includegraphics[width=\textwidth]{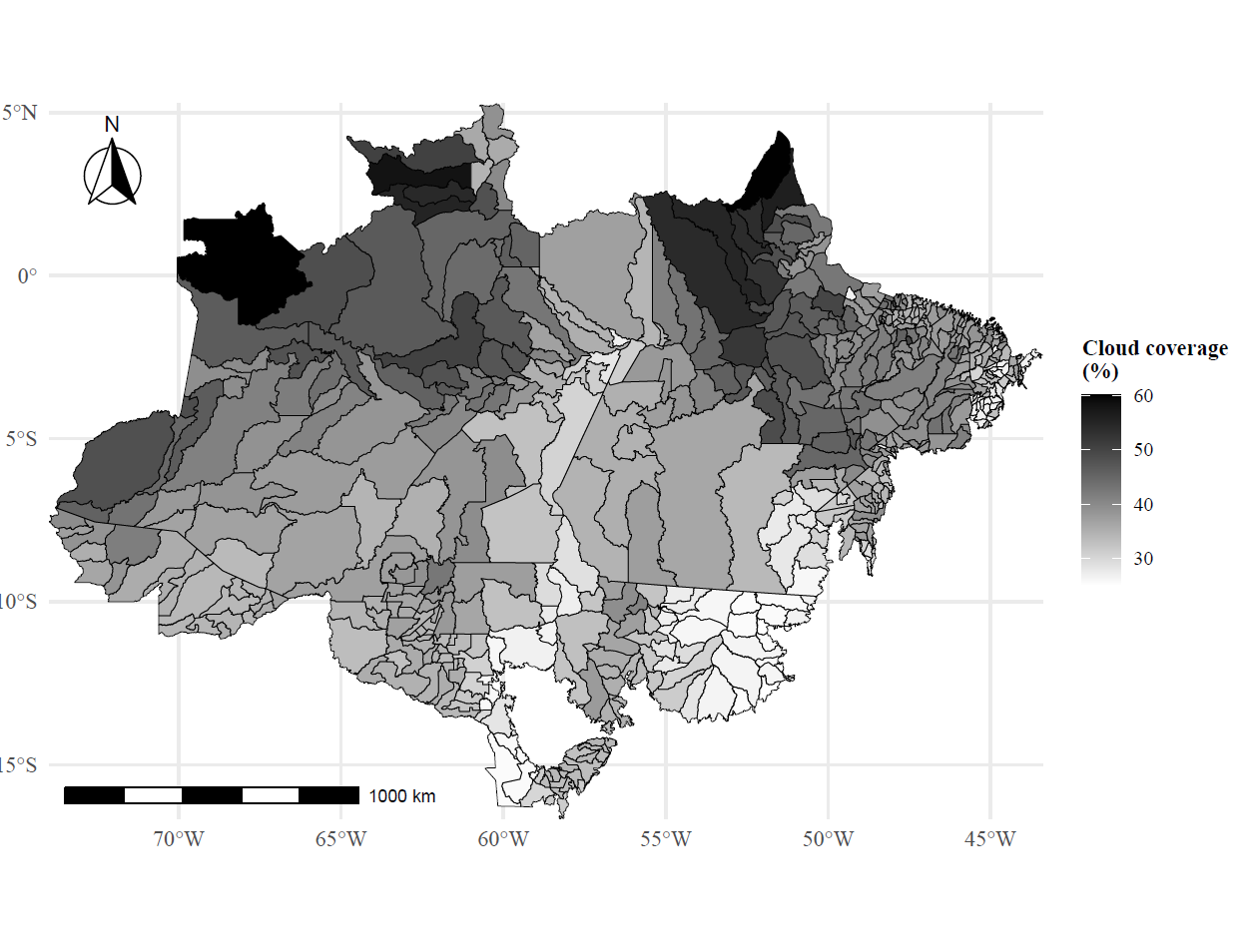}
		\caption{DETER Cloud Coverage}
		\label{fig:deter_2006}
	\end{subfigure}

	\resizebox{\linewidth}{!}{%
		\parbox{\linewidth}{\footnotesize
			\textit{Notes:} Variable labels, units, and sources are as follows. Homicide Rate: homicide per 100,000 inhabitants, Mortality Information System (SIM-DataSUS) and Brazilian Institute for Geography and Statistics (IBGE); Environmental Fines: number of deforestation-fines, Brazilian Institute for the Environment and Renewable Natural Resources (IBAMA); DETER cloud coverage: ratio of cloud to municipal area, Real-Time System for Detection of Deforestation (DETER) from the Brazilian Institute for Space Research (INPE). See Section~\ref{SecData} for details on variable construction.
		}
	}
\end{figure}

Figure~\ref{fig:maps_2006} illustrates the geographical distribution of key variables in 2006, the first year of the sample period. While this snapshot does not aim to represent trends over time, it helps visualize the type of cross-sectional variation present in the data. Panel~\ref{fig:homicide_2006} displays the homicide rate per 100,000 inhabitants, revealing visible heterogeneity across municipalities. Panel~\ref{fig:enforcement_2006} shows the distribution of deforestation-related fines, highlighting spatial differences in environmental enforcement. Panel~\ref{fig:deter_2006} presents DETER cloud coverage — the instrumental variable used in the analysis — which also varies substantially across space.
These maps highlight the considerable spatial heterogeneity in the outcome, treatment, and instrument variables.

To complement the geographical distribution shown in Figure~\ref{fig:maps_2006}, Table~\ref{tab:descriptive_statistics} reports summary statistics for the main variables used in the empirical analysis. Homicide counts and rates exhibit substantial heterogeneity across municipalities and years, with an upward trend throughout the sample period. The average homicide rate rose from 22.9 per 100,000 inhabitants in 2006 to 34.9 in 2016, while the absolute number of homicides nearly doubled over the same period. Environmental enforcement, proxied by the number of deforestation-related fines, peaked in 2008 and declined steadily thereafter. Cloud coverage also fluctuates over time, with notably high levels in 2008 and 2010, potentially impairing satellite-based monitoring and enforcement. These descriptive patterns motivate the instrumental variable strategy.

Satellite, weather, and socioeconomic controls also display considerable variation across municipalities and over time. Among the satellite-based measures, DETER cloud coverage averaged 46\% across the sample, with peaks in 2008 and 2010, while the PRODES system reported large swings in non-observable areas, especially in the late 2000s. Weather conditions remained relatively stable, with average annual precipitation around 6,960 mm and mean temperatures close to 26°C. Socioeconomic indicators evolved gradually: the commodity price index increased over the sample period, the municipal GDP showed consistent growth, population density rose modestly, and education outcomes improved steadily. These patterns highlight relevant sources of heterogeneity in local conditions that are accounted for in the empirical analysis.

\section{Empirical Strategy}\label{SecEmpirical}
This section outlines the empirical strategy used to estimate the causal effect of environmental enforcement on violent crime in the Brazilian Amazon. The analysis focuses on municipal homicide rates as the outcome of interest and uses the number of deforestation-related fines issued under the DETER system as a proxy for enforcement intensity.

The main empirical challenge is that environmental law enforcement may be endogenous to local violence. Municipalities with higher crime rates might receive more enforcement resources, either in response to deteriorating conditions or due to institutional targeting. Additionally, unobserved local characteristics --- such as institutional presence, land tenure patterns, or illegal economic activity --- may simultaneously influence both enforcement and violent crime. In this context, estimating the causal effect of enforcement through ordinary least squares would likely produce biased estimates due to reverse causality and omitted variable bias.

To address potential endogeneity between environmental enforcement and violence, we follow the strategy proposed by \citet{assuncao2023deter} and implement a two-stage least squares instrumental variable approach. The key idea is to isolate exogenous variation in enforcement intensity generated by cloud cover, which obstructs DETER’s real-time satellite monitoring capacity. Because DETER alerts are the primary operational input triggering inspections and sanctions, greater cloud coverage mechanically reduces enforcement. As cloudiness is driven by atmospheric conditions unrelated to local criminal dynamics after controlling for covariates such as rain, temperature and municipality fixed effects, it provides a credible source of exogenous variation for causal identification.

For this strategy to be valid, two conditions must be satisfied. First, cloud coverage must be strongly correlated with enforcement intensity. This is a testable condition and is likely to hold mechanically, since DETER relies on daily optical satellite imagery to detect illegal deforestation. Second, the exclusion restriction must hold: conditional on controls and fixed effects, cloud coverage must affect homicide rates only through its impact on environmental enforcement. While it is unlikely that cloudiness directly influences violent behavior, persistent cloud cover may correlate with structural differences across municipalities, such as remoteness, ecological conditions, or the prevalence of illegal extractive activities. To mitigate this concern, we include a rich set of time-varying controls for satellite observability, weather patterns, and socioeconomic conditions, as well as municipality and year fixed effects. These adjustments help ensure that the identifying variation reflects temporary, exogenous shocks to enforcement capacity rather than deeper institutional or geographic traits.\footnote{In our model from Section \ref{SecModel}, cloud cover represents an exogenous shift in the parameter $\theta$ (the probability of detecting deforestation). While detection may be affected by deforestation itself, cloud cover serves as a valid instrument because---as argued in \cite{assuncao2023deter}---its year-to-year variation, conditional on climate controls such as rainfall and temperature, affects only the detection probability.}\footnote{An important distinction between our specification and that of \cite{assuncao2023deter} is that when the outcome is deforestation, cloud cover could in principle affect the measurement of deforestation directly, because satellites cannot observe land use under heavy cloud cover. This concern is largely absent when the outcome is homicides, which are measured using administrative crime records.} We also formally assess the plausibility of the exclusion restriction using the sensitivity analysis framework proposed by \citet{cinelli2025omitted} in Section \ref{sec:robustness_instrument_validity}.

Under these assumptions, we estimate the causal effect of enforcement on violent crime using the 2SLS specification. The second stage is given by:
\begin{equation}
	Homicide\_Rate_{i,t} = \delta Enforcement_{i,t-1} + \sum_k \theta_k \mathbf{X}_{i,t-1} + \psi_i + \lambda_t + \varepsilon_{i,t}
	\label{eq:second_stage}
\end{equation}
where $Homicide\_Rate_{i,t}$ is the homicide per 100,000 inhabitants of municipality $i$ in year $t$; $Enforcement_{i,t-1}$ is the  the total number of deforestation-related fines issued in municipality $i$ in year $t-1$ and is instrumented by $CloudCover_{i,t-1}$; $\mathbf{X}_{i,t-1}$ is a vector of $k$ municipality-level controls that includes satellite visibility, weather patterns and socioeconomic factors; $\psi_i$ is the municipality fixed effect; $\lambda_t$ is the year fixed effect; and $\varepsilon_{i,t}$ is the idiosyncratic error. Standard errors are clustered at the municipality level.

The enforcement variable is lagged by one year to reflect the documented delay between changes in monitoring capacity and behavioral responses by illegal actors \citep{levitt1997impact, chalfin2017criminal}. This lag structure is consistent with the mechanism through which environmental enforcement may affect violence in the Amazon. In this context, law enforcement targets illegal land-use activities such as unauthorized deforestation, land grabbing, or unlicensed extractive operations. These activities are often associated with criminal networks that use coercion, threats, or violence to secure control over land and resources. When enforcement intensifies, the expected cost of engaging in these activities increases. Over time, this may deter criminal actors, reduce conflict over land, and weaken the economic incentives of violence-driven land exploitation. However, these responses are not immediate because illegal operators may take time to perceive changes in enforcement patterns, reassess risks, or relocate their activities. By introducing a one-year lag, the model captures these dynamics more realistically. Additionally, using a lagged treatment mitigates concerns about reverse causality --- namely, that spikes in violence might trigger greater enforcement in the same year.

To estimate Equation~\eqref{eq:second_stage}, we instrument the enforcement variable using cloud coverage detected by the DETER system. The first stage of the 2SLS approach captures how adverse monitoring conditions affect the issuance of deforestation-related fines. The first-stage specification is given by:
\begin{equation}
	Enforcement_{i,t-1} = \beta CloudCover_{i,t-1} + \sum_k \gamma_k \mathbf{X}_{i,t-1} + \alpha_i + \phi_t + \epsilon_{i,t}
	\label{eq:first_stage}
\end{equation}

\noindent
where $Enforcement_{i,t-1}$ is proxied by the total number of deforestation-related fines issued in municipality $i$ in year $t-1$; $CloudCover_{i,t-1}$ is the average cloud coverage in municipality $i$ in year $t$ measured by DETER; $\mathbf{X}_{i,t-1}$ is a vector of $k$ municipality-level controls that includes satellite visibility, weather patterns and socioeconomic factors; $\alpha_i$ is the municipality fixed effect; $\phi_t$ is the year fixed effect; and $\epsilon_{i,t}$ is the idiosyncratic error. Standard errors are clustered at the municipality level.

\section{Results}\label{SecResults}

Section \ref{SecMainResults} presents our main estimates of the causal effect of environmental enforcement on the homicide rate. Moreover, Section \ref{sec:money} provides a cost-benefit analysis of the DETER program based on the effect found in Section \ref{SecMainResults}. Section \ref{SecFirstStageResults} presents the results of the first-stage regression (Equation \eqref{eq:first_stage}). Section \ref{SecResults_mechanisms} discusses the mechanisms that may explain the effects in the outcome equation. Additionally, Appendix~\ref{app_sec:sex_race} provides heterogeneity analyses by sex, race and age group, showing that the reduction in homicides is concentrated on black male adults.

\subsection{Main Results}\label{SecMainResults}

Table \ref{tab:main_results} presents the main results from the 2SLS estimates of the effect of environmental enforcement on municipal homicide rates, using cloud coverage recorded by the DETER satellite system as an instrument. Panel A reports the second-stage coefficients for the impact of law enforcement on homicide rates (Equation \eqref{eq:second_stage}), while Panel B shows the corresponding first-stage estimates of the relationship between cloud coverage and environmental enforcement (Equation \eqref{eq:first_stage}). The dependent variable is the homicide rate (homicides per 100,000 inhabitants), and the enforcement proxy is the number of deforestation-related fines issued by IBAMA in the previous year.

Columns (1) to (3) in Table~\ref{tab:main_results} report progressively saturated specifications, designed to assess the robustness of the estimated effect to the addition of control variables. The specification in Column (1) includes only satellite-based controls, capturing potential measurement limitations in forest monitoring that could confound the enforcement proxy. Column (2) adds weather variables, which account for climatic conditions that may affect both enforcement capacity and local dynamics in the murder rate. Column (3) introduces a richer set of socioeconomic controls --- such as commodity prices, economic output, population density, and education quality --- to capture municipal-level heterogeneity in crime incentives and institutional presence.

\begin{table}[!htb]
	\begin{center}
		\caption{2SLS --- Law Enforcement and Homicide Rate}
		\label{tab:main_results}
		\begin{tabular}{lccc}
			\toprule
			\toprule

			\multicolumn{4}{c}{{Panel A: Second Stage (Homicide Rate)}} \\
			& (1) & (2) & (3) \\

			\midrule

			Lagged Enforcement & -0.894$^{**}$  & -0.947$^{**}$  & -0.728$^{**}$  \\
			& (0.443)        & (0.452)        & (0.351) \\

			\multicolumn{4}{l}{Average homicide rate across municipalities = 28.16} \\

			& & & \\

			\toprule

			\multicolumn{4}{c}{{Panel B: First Stage (Environmental Enforcement)}} \\
			& (1) & (2) & (3) \\

			\midrule

			DETER Cloud Coverage & -7.495$^{***}$  & -7.436$^{***}$  & -8.880$^{***}$  \\
			& (2.251)         & (2.225)         & (2.192) \\

			& & & \\

			\midrule

			First Stage F-statistic & 11.08 & 11.17 & 16.40 \\

			FE: Municipality \& Year & \checkmark & \checkmark & \checkmark \\

			Controls: & & & \\
			\hspace{1em} Satellite & \checkmark & \checkmark & \checkmark \\
			\hspace{1em} Weather & \xmark & \checkmark & \checkmark \\
			\hspace{1em} Socioeconomic & \xmark & \xmark & \checkmark \\

			Observations & 5,210 & 5,210 & 5,210 \\
			Municipalities & 521 & 521 & 521 \\

			\bottomrule

		\end{tabular}
	\end{center}
	\footnotesize{Notes: 2SLS coefficients are estimated based on Equation \eqref{eq:second_stage} from Section \ref{SecEmpirical}. Panel A presents second-stage estimated coefficients (Equation \eqref{eq:second_stage}). Panel B presents first-stage estimated coefficients (Equation \eqref{eq:first_stage}). Columns (1) to (3) report progressively saturated specifications. Column (1) includes only satellite-based controls: PRODES cloud coverage and non-observable. Column (2) adds weather variables: precipitation and temperature. Column (3) further includes socioeconomic controls: commodity index, GDP, population density, and Ideb scores. ``Homicide Rate'' is the number of homicides per 100,000 inhabitants. ``Lagged Enforcement'' refers to the total number of fines issued and serves as a proxy for environmental enforcement. The dataset is a municipality-by-year panel covering the period 2006-2016. The sample includes all municipalities in the Amazon biome that exhibited variation in forest cover during the sample period and for which deforestation data are available. Standard errors are clustered at the municipality level and reported in parentheses. Significance: $^{*}p<0.1$, $^{**}p<0.05$, $^{***}p<0.01$.}
\end{table}

Across all specifications in Panel A of Table \ref{tab:main_results}, the results point to a negative and statistically significant relationship between environmental enforcement and homicide rates. In Column (1), the estimated coefficient equals –0.89, significant at the 5\% level, indicating that an increase of one fine in the previous year reduces 0.89 homicides per 100,000 inhabitants in the current year. Adding weather controls in Column (2) leaves the estimated coefficient and its significance virtually unchanged. Lastly, after adding socioeconomic covariates, Column (3) reports a smaller estimated coefficient of –0.73, which retains its statistical significance. This specification provides the most credible estimate, as it adjusts for a comprehensive set of observable factors that may influence both enforcement activity and violence.

Taken together, these results suggest that stronger environmental enforcement --- proxied by the issuance of deforestation-related fines --- leads to a sizeable reduction in violent crime. The consistency across specifications strengthens the case for a causal interpretation, provided the instrument is valid.\footnote{In Section \ref{sec:robustness_instrument_validity}, we argue that any omitted variable would require an implausibly large association with the instrument and the outcome to explain away the estimated effect of environmental enforcement on violence.}

To interpret the magnitude of the effect, the main specification in Column (3) shows that the issuance of one additional deforestation-related fine in the previous year reduces the homicide rate by $0.728$ in the current year. Relative to the sample mean of the homicide rate (28.16), this effect corresponds to a 2.58\% decline in the average homicide rate per additional fine issued. To better illustrate the scale of this effect, consider a shift from the 25th to the 75th percentile in the enforcement distribution, i.e., an increase of approximately 8 fines per year. This change is estimated to reduce the homicide rate by 5.82, representing a 20.7\% decline relative to the sample mean.

These findings underscore that, in regions characterized by weak institutions and contested land use, enhancing the state's capacity to monitor and sanction illegal deforestation has a significant deterrent effect on local violence. Consequently, enforcement mechanisms aimed at environmental regulation can also function as tools for reducing conflict in regions where illegal land use and organized crime often overlap.

\subsection{Valuing Violence Reduction}
\label{sec:money}

Our estimates in Section \ref{SecMainResults} imply that the expansion of state presence through DETER prevented approximately 1,477 homicides per year.\footnote{$\text{``Effect of } 0.728 \text{ homicide per 100,000 inhabitants per fine} \times 9.87 \text{ fines for the average municipality} \times 0.3948 \text{ inhabitants (measured in 100,000 people) in the average municipality''}$ gives the average reduction in homicides per municipality. Multiplying this number by the number of municipalities in the Brazilian Amazon (521) yields a reduction of 1,477 homicides per year. In comparison, the total number of homicides per year in the Amazon is 8,790 on average and the total population in the region is 20.5 million (Table \ref{tab:descriptive_statistics}).} This result represents a 15\% reduction in the regional homicides. For comparison, the literature evaluating policies that increase state presence in areas with low state capacity frequently finds large effects on homicides. For example, \citet{Ferraz2023} estimate that retaking control of slums dominated by drug gangs in the city of Rio de Janeiro, Brazil, decreases homicides by 30\% around pacified slums, and \citet{Mancha2025} estimate that the creation of militarized, motorcycle-based police squads in a poor state in Brazil (Ceará) decreases homicides by 57\%.\footnote{Additionally, \cite{chimeli2017use} find that the criminalization of the mahogany trade in the Amazon resulted in an additional 345 homicides per year, considering only the state of Pará (accounting for 38\% of the population in our sample, as of 2010) and covering the period 1999–2013, when homicide rates were lower than in our study period (Figure \ref{fig:homicide_trend}).}

To value the estimated reduction, we use the willingness-to-pay estimates by \cite{dominguez2024willingness}. They estimate a willingness to pay of \$152 per person per year for a 20\% decrease in homicides. Scaling this estimate to our result implies aggregate benefits of roughly \$2.3 billion annually.\footnote{To find this value, multiply \$152 by $\frac{15\%}{20\%}$ and by 20.5 million people.}

For comparison, the combined annual budget of IBAMA and INPE --- a loose upper bound on the costs of DETER according to \cite{assuncao2023deter} --- is about \$622 million. Hence, even abstracting from DETER’s environmental objectives, its law enforcement component alone generates a benefit–cost ratio of at least 3.7.

\subsection{First-Stage Results}\label{SecFirstStageResults}

Panel B of Table~\ref{tab:main_results} reports the estimated first-stage coefficients (Equation \eqref{eq:first_stage}). Cloud coverage observed by DETER is strongly associated with the number of fines issued. In the main specification (Column (3)), the estimated coefficient is –8.88 and statistically significant at the 1\% level, indicating that a one-percentage-point increase in annual cloud cover reduces enforcement intensity by nearly nine fines. The first-stage F-statistic of 16.40 exceeds the conventional threshold of 10 suggested by \citet{stock2002survey}, reinforcing the relevance of cloud coverage as a source of exogenous variation in enforcement intensity.

\subsection{Mechanism Discussion}
\label{SecResults_mechanisms}
This section discusses three plausible channels through which environmental enforcement may reduce violence: (i) a decline in deforestation and related land conflict; (ii) greater state presence and deterrence; and (iii) disruption of illegal economic activities. Findings in the existing literature support these mechanisms, although our analysis is unable to isolate their individual contributions.

First, environmental enforcement reduces deforestation, which can ease disputes over land. \citet{assuncao2023deter} show that the DETER system --- the same monitoring framework used here --- substantially curtailed deforestation in the Amazon biome. The authors found that increasing monitoring and enforcement by half led to an estimated 25\% drop in municipal deforestation. Since land disputes in the region are closely tied to illegal clearing for speculative or productive purposes, curbing deforestation likely eases tensions over land appropriation and undermines a key driver of rural violence. This interpretation aligns with \citet{alston2000land}, who identify land reform struggles and insecure property rights as central sources of violent conflict in the Amazon.

Second, the issuance of environmental fines --- particularly those based on real-time satellite monitoring --- may reinforce the presence of the state in rural areas and enhance crime deterrence. The visible and active role of enforcement agencies can increase the perceived likelihood of surveillance and punishment, not only for environmental crimes but also for other illicit activities. As noted by \citet{de2024environmental}, improved enforcement capacity and stronger property rights reduce the incentives for violent land appropriation, especially in areas where institutional weakness enables armed groups to act with impunity. In this sense, fines serve as credible signals of institutional control.

Third, enforcement actions may disrupt the operation and profitability of illegal economic activities such as land grabbing, unauthorized logging, and unregulated mining. By increasing both the expected costs and logistical challenges of operating outside the law, environmental fines can displace or deter organized groups whose activities often rely on coercion and violence. \citet{chimeli2017use} document the role of violence in sustaining the illegal mahogany trade in the region, while \citet{soares2021ilegalidade} and \citet{pereira2022tale} emphasize the close connection between environmental crimes and local criminal networks. In this view, enforcement shifts the economic incentives of actors who depend on force and impunity to maintain territorial control.\footnote{Regulatory changes in mining or timber extraction could offer an additional explanation for the effect we document. To address this concern, we: (1) re-estimate our main specification including as a control the interaction between annual gold prices from the World Bank and a dummy for municipal mining activity \citep[as in][]{pereira2022tale} constructed using MapBiomas data. The results remain virtually unchanged (point estimate = –0.72, standard error = 0.35 in the specification of Table 1, Column (3), with the gold–mining interaction included); and (2) include as a control the interaction between annual tropical hardwood log prices from the World Bank and a dummy for mahogany presence from \citet{soares2021ilegalidade}. Again, the results are virtually unchanged (point estimate = –0.79, standard error = 0.37 in the specification of Table 1, Column (3), with the timber–mahogany interaction included).}

Taken together, these mechanisms provide a plausible framework for interpreting the estimated reduction in homicides as a downstream effect of environmental enforcement. The deforestation channel is directly supported by prior causal evidence, while the deterrence and disruption effects are consistent with well-documented institutional dynamics in the Amazon. While our empirical strategy does not distinguish among these channels, the literature suggests they are likely to operate simultaneously.

\section{Extensions and Robustness Checks}\label{SecRobChecks}
This section evaluates the robustness of the main results through sensitivity analyses and different outcome definitions. In Section \ref{sec:robustness_instrument_validity}, we examine the validity of the instrument following the sensitivity analysis proposed by \citet{cinelli2025omitted}. In Section \ref{sec:robustness_homicide_classification}, as a robustness check, we re-estimate the model using alternative measures of violence. We first restricted the outcome to ICD-10 categories X85--Y09, which capture assaults committed through various means. Next, we extend the definition to include ICD-10 categories Y10--Y34, which capture events of undetermined intent, and, lastly, we incorporate ICD-10 category Y35, which captures legal intervention. Additionally, in Section \ref{sec:robustness_distributional_regression}, as an extension, we estimate a set of distributional regressions to assess whether the effects of environmental enforcement extend beyond the average homicide rate. Specifically, we examine its impact on two binary outcomes: (i) the incidence of any violence (i.e., a dummy equal to one if the homicide rate is greater than zero), and (ii) the presence of endemic violence (i.e., a dummy equal to one if the homicide rate exceeds 10 homicides per 100,000 inhabitants).

In Appendix \ref{sec:robustness_conservation_policies}, we present an additional robustness check that controls for other conservation policies coexisting with the DETER program.

\subsection{Sensitivity Analysis: Exogeneity and Exclusion Restriction} \label{sec:robustness_instrument_validity}

A credible instrumental variable must be relevant (i.e., strongly correlated with the endogenous regressor), excluded from the outcome equation, and independent from the unobservables that determine the outcome. While the first condition is directly testable through the first-stage regression (Section \ref{SecFirstStageResults}), the last two conditions require indirect tests. One way to assess the plausibility of these two conditions is to use the sensitivity analysis tool proposed by \cite{cinelli2025omitted}.

This tool leverages the property that, in a linear model, the exclusion restriction and the exogeneity assumption can be jointly stated as imposing that the correlation between the instrument and the unobservables that determine the outcome is zero. If this correlation is not zero, then there exists an omitted variable $W$ that creates bias in the 2SLS estimand.

To have a better understanding of this possible bias, \cite{cinelli2025omitted} shows that it depends on two key parameters: (i) the share of the residual variance of the outcome that is explained by the omitted variable $W$ after controlling for all covariates and the instrument $\left( R^2_{Y \sim W \mid Z, \mathbf{X}} \right)$, and (ii) the share of the residual variance of the instrument that is explained by the omitted variable $W$ after controlling for all covariates $\left( R^2_{Z \sim W \mid \mathbf{X}} \right)$. The key idea of this sensitivity analysis tool is to find how large $R^2_{Y \sim W \mid Z, \mathbf{X}}$ and $R^2_{Z \sim W \mid \mathbf{X}}$ must be to imply that the 2SLS estimand is pure bias, i.e., that the true effect is zero and the 2SLS regression captures solely the correlation between the instrument and the unobservables of the outcome equation.

Many values for $R^2_{Y \sim W \mid Z, \mathbf{X}}$ and $R^2_{Z \sim W \mid \mathbf{X}}$ may imply that the 2SLS estimand is pure bias. These values are known as the breakdown frontier. For brevity, \cite{cinelli2025omitted} focus on one value of this frontier: the robustness value $\overline{r}$. This statistic answers the following question: if there is an omitted variable that equally explains the residual outcome variance and the residual instrumental variance (i.e., $R^2_{Y \sim W \mid Z, \mathbf{X}} = R^2_{Z \sim W \mid \mathbf{X}} \eqqcolon \overline{r}$), how large must its explanatory power be to imply that the 2SLS estimand contains only bias?

In our data, the answer to this question is $\overline{r}=$3.35\%. In other words, if there exists an omitted variable $W$ that explains 3.35\% of (i) the residual variance of the outcome $\left( R^2_{Y \sim W \mid Z, \mathbf{X}} \right)$, and (ii) the residual variance of the instrument $\left( R^2_{Z \sim W \mid \mathbf{X}} \right)$, then the true linear effect of environmental enforcement on the homicide rate is zero.

This number raises another question: is the existence of such an omitted variable plausible? To answer this question, we may use the covariates in Equation \eqref{eq:second_stage} to approximate  $R^2_{Y \sim W \mid Z, \mathbf{X}}$ and $R^2_{Z \sim W \mid \mathbf{X}}$. \cite{cinelli2025omitted} propose to omit each covariate $X_{k}$ separately and estimate (i) the share of the residual variance of the outcome that is explained by the omitted covariate $X_{k}$ after controlling for all the other covariates and the instrument $\left( R^2_{Y \sim X_{k} \mid Z, \mathbf{X}_{-k}} \right)$, and (ii) the share of the residual variance of the instrument that is explained by the omitted covariate $X_{k}$ after controlling for all the other covariates $\left( R^2_{Z \sim X_{k} \mid \mathbf{X}_{-k}} \right)$.

In our data, the covariate that explains the largest share of the residual variance of the instrument and the outcome is the Commodity Index. We find that it explains 0.08\% of the residual variance of the outcome (i.e., $R^2_{Y \sim X_{k} \mid Z, \mathbf{X}_{-k}} = 0.0008$) and 1.88\% of the residual variance of the instrument (i.e., $R^2_{Z \sim X_{k} \mid \mathbf{X}_{-k}} = 0.0188$). Since these numbers are much smaller than the robustness value ($\overline{r}=$3.35\%), the existence of an omitted variable capable of explaining away our estimated effect is implausible.

Consequently, the relatively high robustness values observed for DETER coverage suggest that our main results are not easily overturned by confounding factors of similar magnitude to a key observable covariate. These findings enhance the credibility of the IV design by demonstrating that substantial confounding would be required to meaningfully challenge the interpretation of the estimated effects as causal.

\subsection{Alternative Homicide Classification}
\label{sec:robustness_homicide_classification}
In the main specification, the dependent variable includes all deaths classified under ICD-10 codes X85--Y09 (assaults), Y10--Y34 (events of undetermined intent), and X60--X84 (intentional self-harm). This broader definition is inspired by the approach adopted by \cite{ipea2024homicidios,ipea2025} to better capture lethal violence in contexts where underreporting and misclassifications are prevalent.\footnote{Additional robustness checks using other disaggregated homicide categories (firearm assaults, assaults with sharp/blunt objects, and legal intervention) are reported in Appendix~\ref{app_sec:others_ICD}.}

A potential concern is that our inclusive classification may inadvertently be capturing an effect on actual intentional self-harm. If this were the case, it would bias our results against our findings. The most plausible channel linking environmental enforcement to intentional self-harm would operate through income: stricter enforcement reduces deforestation, which may lower earnings in affected areas, potentially increasing intentional self-harm. This mechanism would predict that enforcement raises intentional self-harm events, rather than lowering them, as we find in Table \ref{tab:main_results}. Moreover, in Appendix \ref{app_sec:others_ICD}, we show that the effect of lagged enforcement on intentional self-harm is statistically insignificant and the point estimate is negative.

To further assess the robustness of the results to alternative outcome definitions, Table~\ref{tab:homicide_specification} compares the 2SLS estimates using different homicide classifications. The outcome variable in Column (1) considers only deaths explicitly coded as assaults (ICD-10 X85--Y09) as used by \cite{Dix-Carneiro2018}, while the outcome variable in Column (2) broadens the definition by also incorporating events of undetermined intent (ICD-10 Y10--Y34). Column (3) corresponds to the main specification, and Column (4) adds deaths due to legal intervention (ICD-10 Y35). All regressions include municipality and year fixed effects as well as the full set of controls.

\begin{table}[ht]
	\centering
	\caption{2SLS --- Estimates under Alternative Homicide Classifications}
	\label{tab:homicide_specification}
	\begin{tabularx}{\textwidth}{lCCCC}
		\toprule
		\toprule

		& \multicolumn{4}{c}{Homicide Rate} \\
		& (1) & (2) & (3) & (4) \\

		\midrule

		Lagged Enforcement  & -0.501$^{*}$ & -0.550$^{*}$ & -0.728$^{**}$ & -0.727$^{**}$ \\
		& (0.292)      & (0.309)      & (0.351) & (0.351) \\

		& & &  &\\

		\midrule

		Average homicide rate across municipalities  & 22.35 & 24.30 & 28.16 & 28.20 \\

		& & &  & \\

		\midrule

		FE: Municipality \& Year & \checkmark & \checkmark & \checkmark & \checkmark \\

		& & & & \\

		All controls & \checkmark & \checkmark & \checkmark & \checkmark \\

		& & & & \\

		ICD-10 classification: & & & & \\
		\hspace{1em} X85--Y09 (assaults by various means) & \checkmark & \checkmark & \checkmark & \checkmark \\
		\hspace{1em} Y10--Y34 (events of undetermined intent) & \xmark & \checkmark & \checkmark & \checkmark \\
		\hspace{1em} X60--X84 (intentional self-harm) & \xmark & \xmark & \checkmark & \checkmark \\
		\hspace{1em} Y135 (legal intervention) & \xmark & \xmark & \xmark & \checkmark \\
		& & & \\

		Observations & 5,210 & 5,210 & 5,210 & 5,210 \\
		Municipalities & 521 & 521 & 521 & 521 \\

		\bottomrule

	\end{tabularx}

	\vspace{0.1em}

	\resizebox{\linewidth}{!}{%
		\parbox{\linewidth}{\footnotesize
			\textit{Notes:} 2SLS coefficients are estimated based on Equation~\eqref{eq:second_stage} from Section~\ref{SecEmpirical}. Column (1) includes only assaults by various means (ICD-10 X85--Y09), Column (2) adds events of undetermined intent (ICD-10 Y10--Y34), Column (3) adds intentional self-harm  (ICD-10 X60--X84), and Column (4) adds legal intervention (ICD-10 Y-35). ``Homicide Rate'' is the number of homicides per 100,000 inhabitants. ``Lagged Enforcement'' refers to the total number of fines issued and serves as a proxy for law enforcement effectiveness. The set of control variables contains PRODES cloud coverage and non-observable (satellite); precipitation and temperature (weather); and commodity index, GDP, population density and Ideb scores (socioeconomic). The dataset is a municipality-by-year panel covering the period 2006-2016. The sample includes all Amazon biome municipalities that exhibited variation in forest cover during the sample period and for which deforestation data are available. Standard errors are clustered at the municipality level and reported in parentheses. Significance: $^{*}p<0.1$, $^{**}p<0.05$, $^{***}p<0.01$.
		}
	}

\end{table}

The results remain consistent across all definitions. In Column (1), the estimated effect of enforcement is –0.501 and statistically significant at the 10\% level. This implies that the issuance of one additional deforestation-related fine is associated with a reduction of approximately 0.50 homicides per 100,000 inhabitants, representing a 2.2\% decrease relative to the average homicide rate across municipalities. In Column (2), the estimated coefficient becomes –0.55, also statistically significant at the 10\% level. This effect implies a 2.3\% reduction relative to the average homicide rate across municipalities. In Column (4), the estimated coefficient is -0.73 and remains significant at the 5\% level. Relative to the average homicide rate across municipalities, this effect represents a reduction of 2.6\%. Importantly, the relative effects are approximately 2.2\%--2.6\% in all specifications, and the estimated coefficients are statistically indistinguishable across specifications.

This stability compared to the main specification suggests that specific coding choices do not drive the main findings. Consequently, it reinforces the conclusion that environmental enforcement contributes to reducing lethal violence, even under conservative definitions of homicide.

Moreover, we separately consider the inclusion of deaths classified under ICD-10 code Y35 (legal intervention) in Appendix~\ref{app_sec:others_ICD}. These cases correspond to killings by state agents in situations where the use of lethal force is deemed lawful, such as armed confrontations with the police. While distinct from conventional homicides, they represent a form of lethal violence that may respond to changes in environmental enforcement through increased interaction between law enforcement and illegal activities. We find no statistically significant effect of environmental enforcement on rates of fatalities in confrontations with state agents. This result is consistent with the institutional context in which IBAMA, rather than police forces, carries out most enforcement activities, and confrontations rarely escalate into lethal violence.


\subsection{Violence Incidence and Severity}
\label{sec:robustness_distributional_regression}
To further probe the robustness of the main results, we estimate a set of distributional regressions. Rather than focusing solely on the continuous homicide rate, this approach assesses whether environmental enforcement affects the likelihood that a municipality experiences any violence or presents a pattern of persistently high violence over time. We define ``any violence'' as a binary indicator equal to one if the homicide rate is greater than zero, and endemic violence as a binary indicator equal to one if the homicide rate exceeds 10 homicides per 100,000 inhabitants in a given year. The latter threshold draws from the public health literature, where homicide rates above 10 are often considered a marker of endemic violence by institutions such as the World Health Organization (WHO).

The effects on both dependent variables are estimated using the same 2SLS strategy adopted in the baseline regressions. Table~\ref{tab:distributional_regression} presents the second-stage estimates for the effect of environmental enforcement on the likelihood of ``any violence'' (Column (1)) and endemic violence (Column (2)).

\begin{table}[ht]
	\centering
	\caption{2SLS --- Distributional Regression: Violence and Endemic Violence}
	\label{tab:distributional_regression}
	\begin{tabularx}{\textwidth}{lCC}
		\toprule
		\midrule

		& Murder Rate $> 0$ &  Murder Rate $> 10$\\
		& (1) & (2) \\

		\midrule

		Lagged Enforcement & -0.012$^{*}$  & -0.018$^{**}$  \\
		& (0.006)       & (0.008) \\

		& & \\

		\midrule

		Average of the outcome variable across municipalities  & 0.84 & 0.73 \\

		& &  \\

		\midrule

		FE: Municipality \& Year & \checkmark & \checkmark \\

		& & \\

		All controls & \checkmark & \checkmark \\

		& & \\

		Observations & 5,210 & 5,210 \\
		Municipalities & 521 & 521 \\

		\bottomrule

	\end{tabularx}

	\vspace{0.1em}

	\resizebox{\linewidth}{!}{%
		\parbox{\linewidth}{\footnotesize
			\textit{Notes:} 2SLS coefficients are estimated based on Equation~\eqref{eq:second_stage} from Section~\ref{SecEmpirical}. The outcome variable in Column (1) is a dummy variable indicating whether there was at least one homicide in municipality $i$ in year $t$ (``any violence'').  The outcome variable in Column (2) a dummy variable indicating whether there was at least one homicide in municipality $i$ in year $t$ (``endemic violence''). ``Homicide Rate'' refers to the number of homicides per 100,000 inhabitants. ``Lagged Enforcement'' refers to the total number of fines issued and serves as a proxy for law enforcement. The set of control variables contains PRODES cloud coverage and non-observable (satellite); precipitation and temperature (weather); and commodity index, GDP, population density and Ideb scores (socioeconomic). The dataset is a municipality-by-year panel covering the period 2006-2016. The sample includes all Amazon biome municipalities that exhibited variation in forest cover during the sample period and for which deforestation data are available. Standard errors are clustered at the
			municipality level and reported in parentheses.  Significance: $^{*}p<0.1$, $^{**}p<0.05$, $^{***}p<0.01$.
		}
	}

\end{table}

The results reveal a consistent pattern. In Column (1), the coefficient of -0.012 indicates that each additional environmental fine reduces the probability that a municipality records any homicide by 1.2 percentage points. In Column (2), the effect on endemic violence is even larger in magnitude: each additional fine lowers the probability of recording a homicide rate above 10 by 1.8 percentage points. Both estimates are statistically significant and substantively meaningful. Taken together, these results confirm that the reduction in lethal violence associated with environmental enforcement is not driven solely by marginal shifts in the homicide distribution, affecting both the incidence and severity of violence across municipalities.

\section{Conclusion}\label{SecConclusion}
This study investigates whether environmental enforcement policies generate broader social benefits by reducing violence in the Brazilian Amazon. Exploiting an instrumental variable strategy based on exogenous cloud-induced variation in satellite visibility, we identify the causal impact of enforcement intensity, proxied by deforestation-related fines, on municipal homicide rates. The empirical analysis demonstrates that stronger environmental enforcement significantly reduces homicide rates. The results indicate that each additional fine issued by enforcement agencies corresponds to a decrease of approximately 0.78 homicides per 100,000 inhabitants, representing a meaningful 2.6\% reduction relative to the average homicide rate. Alternatively, an increase from the 25th to the 75th percentile in the distribution of enforcement intensity corresponds to a reduction of approximately 20.7\% in homicide rates. Robustness checks --- alternative violence definitions, sensitivity analyses for omitted variable bias, and distributional regressions on violence incidence --- further validate the credibility of these findings.

The implications of these findings are significant for policymakers addressing governance challenges in environmentally sensitive areas. Traditionally, policy discussions around conservation and development have been framed as involving trade-offs, suggesting that strengthened environmental regulation could constraint economic opportunities and exacerbate social tensions. However, our results challenge this conventional wisdom by demonstrating that command-and-control policies designed to prevent illegal deforestation can simultaneously mitigate violent crime. Consequently, policymakers should integrate environmental enforcement measures into broader public safety and governance strategies, especially in regions characterized by institutional fragility and limited accessibility. Importantly, satellite monitoring systems, such as the one we evaluate in this paper, can be replicated across many areas in the developing world due to their relatively low cost.

As with any research, ours is subject to limitations that future studies could address. First, the current analysis does not explicitly investigate the underlying political economy factors influencing enforcement allocation or the strategic behavior of criminal actors responding to enforcement efforts. Understanding how local political dynamics, economic interests, and criminal networks interact with enforcement practices remains crucial for designing contextually effective policies.\footnote{The work of \cite{Dahis2025} is a first step in this direction.} Second, this study focuses on the immediate effects of environmental law enforcement on homicides. However, such interventions are likely to have long-run consequences, including intergenerational effects. Investigating these longer-term impacts is an important avenue for future research.

\singlespace

\bibliography{references_part2}


\pagebreak

\newpage

\pagebreak

\setcounter{table}{0}
\renewcommand\thetable{A.\arabic{table}}

\setcounter{figure}{0}
\renewcommand\thefigure{A.\arabic{figure}}

\setcounter{equation}{0}
\renewcommand\theequation{A.\arabic{equation}}

\appendix

\begin{center}
	\huge
	Supporting Information

	(Online Appendix)

\end{center}

\doublespacing
\normalsize

\section{Detailed Institutional Background: Crime, Environment, and Governance} \label{AppBack}

\setcounter{table}{0}
\renewcommand\thetable{A.\arabic{table}}

\setcounter{figure}{0}
\renewcommand\thefigure{A.\arabic{figure}}

\setcounter{equation}{0}
\renewcommand\theequation{A.\arabic{equation}}

\setcounter{theorem}{0}
\renewcommand\thetheorem{A.\arabic{theorem}}

\setcounter{proposition}{0}
\renewcommand\theproposition{A.\arabic{proposition}}

\setcounter{corollary}{0}
\renewcommand\thecorollary{A.\arabic{corollary}}

\setcounter{assumption}{0}
\renewcommand\theassumption{A.\arabic{assumption}}

\setcounter{definition}{0}
\renewcommand\thedefinition{A.\arabic{definition}}

\setcounter{Lemma}{0}
\renewcommand\theLemma{A.\arabic{Lemma}}

This section provides a more detailed account of the issues introduced in Section~\ref{SecBack}. We begin by discussing the economic theory of crime. Building on this theoretical foundation, we then examine the escalation of violence in the Brazilian Amazon, with a focus on its links to illegal land use, natural resource exploitation, and conflicts over territorial control. Next, we turn to patterns of deforestation, often driven by the same illicit dynamics and reinforced by institutional gaps. Finally, we describe the environmental enforcement strategies implemented by the Brazilian government to curb deforestation, focusing in the satellite-based monitoring program.

\subsection{Violence in the Amazon}

Crime is not a random or irrational phenomenon. Economic models of criminal behavior emphasize that individuals respond to incentives and constraints, weighing the expected benefits of illegal activities against the probability and severity of punishment \citep{becker1968crime,ehrlich1973participation}. In this framework, criminal actions arise when the expected utility of engaging in crime exceeds that of legal alternatives. This logic has informed a vast empirical literature showing that crime rates are sensitive to changes in enforcement, economic conditions, and institutional environments \citep{levitt1997impact, lochner2004effect}. More recent work extends these insights to broader political economy settings, showing that weak institutions, low state capacity, and contested property rights can systematically undermine deterrence and foster environments conducive to violence and predation \citep{soares2004development, acemoglu2005institutions}.

From this perspective, violence is not merely a symptom of social disintegration but a rational response to institutional voids and strategic opportunities. In settings where the rule of law is fragile and state presence limited, illegal markets may rely on coercion to enforce transactions, resolve disputes, and control territory \citep{draca2011panic, glaeser1996crime}. In particular, resource-rich areas with unclear property rights and weak enforcement are especially vulnerable to conflict, as competing actors use violence as a tool to appropriate rents and deter rivals. These underlying mechanisms are especially relevant in frontier regions, where governance often lags behind economic exploitation and organized crime can substitute for formal institutions.

These dynamics are not merely theoretical. They find expression in the Brazilian Amazon, where the institutional conditions described above have created an environment highly conducive to organized violence and criminal appropriation. Violence in the region has distinct and troubling characteristics \citep{fbsp2021amazon}. Unlike most other regions of the country, the Amazon has experienced a persistent increase in homicides over the past decade. While national trends show a general decline in lethal violence, the North region, which comprise much of the Amazon biome, has followed the opposite trajectory \citep{ipea2025}. This pattern reflects a combination of institutional fragility, conflicts over land and natural resources, and the growing influence of organized crime networks operating in remote areas.

A particularly striking feature of violence in the region is its rural concentration. In contrast to the urban character of most violent crime in Brazil, the majority of homicides in the Amazon occur in rural settings, often near deforestation frontiers or contested landholdings \citep{fearnside2008roles}. The advancement of the agricultural frontier, typically under weak rule of law, intensifies disputes over land, fosters land grabbing, and creates opportunities for illegal logging and mining activities --- all of which are frequently associated with coercion and violence \citep{hrw2019}.

Empirical studies have shown that violence can be instrumental in facilitating illegal market transactions, particularly in contexts where property rights are poorly defined and enforcement is limited \citep{chimeli2017use, de2024environmental}. In the Amazon, criminal groups have exploited these institutional gaps to assert territorial control and secure access to valuable natural resources. As \citet{soares2021ilegalidade} argue, the expansion of illicit activities such as illegal logging, gold mining, and land speculation has gone hand in hand with increased violence, especially in municipalities with weak local institutions and low state presence.

These patterns are also reflected in the data used in this study. Figure~\ref{fig:homicide_trend} shows the evolution of homicide rates
in municipalities located inside and outside the Amazon biome between 2000 and 2016.\footnote{Section~\ref{SecData} details on the construction of the homicide rate variable.}  At the beginning of the period, municipalities outside the biome had higher rates of lethal violence. However, trajectories diverged in the early 2000s, as homicide rates in Amazon municipalities began to rise more rapidly, while rates in non-Amazon areas remained relatively stable. By 2016, this divergence had become substantial: average homicide rates in the Amazon far exceeded those in the rest of the country. This trend reinforces the idea that institutional fragility and the expansion of illegal markets have created a distinct environment marked by growing and persistent violence in the region.

Between 2006 and 2016 --- the core period of this study --- the average homicide rate in Amazon municipalities rose from 33.1 to 52.1 per 100,000 inhabitants, a 57.3\% increase. In contrast, non-Amazon municipalities experienced only a modest rise of 8.0\%, from 36.8 to 39.8. To ensure that the striking pattern of rising and comparatively higher homicide rates in the Amazon is not merely the result of aggregation masking regional heterogeneity, we also disaggregated the data by Brazil’s official geographic regions. This additional breakdown further supports the main result. As shown in Table~\ref{tab:homicide_regions}, the North region recorded the sharpest increase in homicide rates over the period, rising from 33.1 in 2006 to 53.5 in 2016, a 61.5\% growth.\footnote{
	The slight difference between this two growth rates reflects the fact that not all municipalities in the North region are part of the Amazon biome, and not all Amazon biome municipalities belong to the North region.
}
Together, the figure and the table provide empirical confirmation of the institutional dynamics discussed above, highlighting the concentration of violence in the Amazon region.

\begin{table}[ht]
	\centering
	\caption{Homicide Rates and Growth by Region (2006–2016)}
	\label{tab:homicide_regions}
	\resizebox{\textwidth}{!}{%
		\begin{tabular}{lccccccccccc|cccccccccc}

			\toprule
			\midrule

			\textbf{Region} & \multicolumn{11}{c|}{\textbf{Homicide Rate (per 100,000)}} & \multicolumn{10}{c}{\textbf{\% Growth Relative to 2006}} \\
			\cmidrule(lr){2-12} \cmidrule(lr){13-22}
			& 2006 & 2007 & 2008 & 2009 & 2010 & 2011 & 2012 & 2013 & 2014 & 2015 & 2016
			& 2007 & 2008 & 2009 & 2010 & 2011 & 2012 & 2013 & 2014 & 2015 & 2016 \\
			\midrule

			North & 33.1 & 34.9 & 38.1 & 41.6 & 42.2 & 42.0 & 42.4 & 42.7 & 43.4 & 49.3 & 53.5 & 5.4\% & 15.1\% & 25.7\% & 27.5\% & 26.9\% & 28.1\% & 29.0\% & 31.1\% & 48.9\% & 61.6\% \\

			\midrule

			Northeast & 38.9 & 41.6 & 43.4 & 46.4 & 45.3 & 47.9 & 48.5 & 51.0 & 50.6 & 53.3 & 56.5 & 6.9\% & 11.6\% & 19.3\% & 16.5\% & 23.1\% & 24.7\% & 31.1\% & 30.1\% & 37.0\% & 45.2\% \\

			\midrule

			Center-West & 37.3 & 37.2 & 41.6 & 40.2 & 39.9 & 43.3 & 46.5 & 46.8 & 44.7 & 44.9 & 42.6 & -0.3\% & 11.5\% & 7.8\% & 7.0\% & 16.1\% & 24.7\% & 25.5\% & 19.8\% & 20.4\% & 14.2\% \\

			\midrule

			South & 33.2 & 33.6 & 35.4 & 36.2 & 33.4 & 34.3 & 32.9 & 32.5 & 32.9 & 35.2 & 35.4 & 1.2\% & 6.6\% & 9.0\% & 0.6\% & 3.3\% & -0.9\% & -2.1\% & -0.9\% & 6.0\% & 6.6\% \\

			\midrule

			Southeast & 36.5 & 35.0 & 34.9 & 33.5 & 31.4 & 31.3 & 31.4 & 31.7 & 29.7 & 29.1 & 30.1 & -4.1\% & -4.4\% & -8.2\% & -14.0\% & -14.2\% & -14.0\% & -13.2\% & -18.6\% & -20.3\% & -17.5\% \\

			\midrule
			\bottomrule
		\end{tabular}%
	}

	\vspace{0.5em}

	\resizebox{\linewidth}{!}{%
		\parbox{\linewidth}{\tiny
			\textit{Notes:} This table presents average homicide rates by region from 2006 to 2016 (left panel), and cumulative percentage growth relative to 2006 levels (right panel). See Section~\ref{SecData} for details on variable construction.
		}
	}
\end{table}

\subsection{Deforestation: Dynamics and Drivers}

The Brazilian Amazon is the largest continuous tropical forest in the world, covering approximately 4.2 million square kilometers --- nearly 50\% of Brazil’s national territory. Of this total, about 2.2 million square kilometers are protected as indigenous lands or conservation units and an estimated area of 700,000 square kilometers consist of undesignated public forests, where any clearing is deemed illegal \citep{azevedo2018no, gandour2018forest}. In the early 2000s, Brazil distinguished itself as the country responsible for the largest amount of tropical forest loss, both in absolute numbers and as a proportion of its forested area \citep{ipcc2007synthesis}. By 2025, roughly 700,000 square kilometers (more than 15\%) of the Amazon’s original forest cover had already been cleared  \citep{inpeTerrabrasilis}.

Most of the deforestation observed in the Brazilian Amazon over the past two decades has been illegal \citep{assuncao2023deter}. Under the Brazilian Forest Code,\footnote{
	The Brazilian Forest Code is a set of federal laws (primarily Law No. 12.651/2012) regulating the use and protection of native vegetation on private rural properties in Brazil.
} legal deforestation is permitted only within defined limits and under strict regulation --- for example, inside rural private properties and with formal authorization. However, even in cases where deforestation could be legal in principle, property-level assessments reveal widespread noncompliance with environmental laws \citep{godar2012responsible}. Moreover, a significant portion of forest clearing occurs in areas where all deforestation is prohibited, such as protected lands, indigenous territories, and state-owned undesignated forests \citep{moutinho2023untitled}.  In these cases, deforestation frequently serves as a strategy to assert possession, particularly in undesignated lands, where demonstrating productive use can strengthen claims for future land titles \citep{aldrich2012contentious, brown2016land}.

These patterns reflect deeper institutional and economic dynamics. Deforestation accelerated during the mid-20th century, following the construction of the nation’s capital  Brasília and federal roadways that enabled colonization and speculative land occupation in remote forest areas \citep{fearnside2005deforestation}. A second wave in the 1990s was driven by rising global demand for commodities such as beef and soy \citep{morton2006cropland}. During both periods, public policy incentivized frontier expansion without resolving overlapping land claims, contributing to legal uncertainty and conflict. Recent evidence shows that most forest loss occurs on undesignated or disputed lands, where legal ownership is contested among formal owners, squatters, and land grabbers \citep{moutinho2023untitled}. Despite several regularization initiatives, more than half of the Legal Amazon still faced unresolved land tenure by the mid-2000s \citep{barreto2008dono}.

Deforestation in the Amazon is both widespread and highly concentrated. A small number of farms account for a disproportionate share of forest clearing \citep{rajao2022risk}. The main proximate drivers are agricultural conversion and illegal land grabbing, which often form part of broader strategies of territorial appropriation. These strategies rely not only on environmental degradation but also on coercive practices. In regions where enforcement is weak and property rights are uncertain, deforestation and violence frequently operate together as tools to consolidate claims or pressure the state to recognize informal occupations \citep{alston1999model, alston2000land}. This interaction is consistent with broader findings that illegal economic activities --- such as land grabbing and mining --- are systematically associated with local conflict and violence \citep{angrist2008rural, idrobo2014illegal}.

These deforestation dynamics have raised increasing concern at both national and international levels. The Amazon plays a critical role in stabilizing the global climate, and continued forest loss threatens to undermine climate mitigation efforts worldwide \citep{ipcc2023syr}. In response to growing environmental and diplomatic pressure, the Brazilian government has implemented a series of monitoring and enforcement mechanisms aimed at curbing illegal deforestation.

\subsection{Environmental Law Enforcement and Monitoring}

Environmental law enforcement in Brazil is supported by a complex institutional framework involving multiple actors. The federal government, through the Ministry of the Environment, oversees the implementation of environmental policy. The Brazilian Institute for the Environment and Renewable Natural Resources (IBAMA) is the main enforcement agency responsible for monitoring compliance, conducting inspections, and issuing sanctions. Complementing IBAMA’s efforts, the Chico Mendes Institute for Biodiversity Conservation (ICMBio) manages protected areas, while state-level agencies perform similar tasks within their jurisdictions.

Until the early 2000s, enforcement operations were mostly reactive and relied on in-person field inspections with limited spatial and temporal coverage. A major institutional shift occurred in 2004 with the launch of the Action Plan for the Prevention and Control of Deforestation in the Legal Amazon (Plano de Prevenção e Controle do Desmatamento na Amazônia Legal, or PPCDAm). The PPCDAm aimed to reduce deforestation systematically and to promote a sustainable development model for the Amazon. It adopted an integrated approach, combining land-use planning, supply chain regulation, environmental enforcement, and technological innovation. A key component of this strategy was the introduction of real-time remote sensing tools to support enforcement agencies.

At the core of the PPCDAm is the satellite-based Real-Time Detection of Deforestation System (DETER), developed and operated by INPE. The DETER system processes satellite imagery to rapidly identify signs of forest cover change, issuing georeferenced alerts that guide inspection teams in the field. Between 2004 and 2017, DETER issued more than 70,000 alerts, corresponding to approximately 88,000 km² of forest disturbance \citep{inpe_deter_modis_2004_2017}. DETER initially relied on MODIS sensor data from NASA’s Terra satellite, with a spatial resolution of 250 meters and daily coverage. This configuration allowed for frequent updates, though it limited detection to forest changes larger than 25 hectares. Analysts at INPE visually interpret imagery using spectral mixture models and predefined classification rules, identifying various categories of disturbance: clear-cut deforestation (with or without remaining vegetation), forest degradation (including fire scars), and logging activity.\footnote{
	Detailed methodological documentation is provided in \citet{inpe_deter_methodology_2022}.
}

Despite its technological advantages, enforcement remains challenging in practice. First, the alerts may reflect forest loss that occurred earlier but was only detected later due to cloud cover. Cloud-related blind spots are particularly relevant in empirical applications. Because enforcement depends on visibility, DETER-generated alerts effectively increase the likelihood of inspection where and when cloud cover is low. Second, in a setting marked by insecure property rights and limited local state presence, identifying and locating offenders is far from straightforward \citep{alston2000land, mueller2018property}. Even when responsible parties are identified, applying the necessary sanctions often requires catching offenders in the act. In this context, the real-time alerts generated by DETER play a critical role. By increasing the likelihood of timely inspections, these alerts raise the perceived probability of detection, enhancing deterrence \citep{ferreira2023satellites}.

Upon receiving an alert, IBAMA may deploy teams to inspect the area, and when infractions are confirmed, issue fines, apply embargoes, and/or seize equipment. In general, deforestation-related fines are often used as stand-alone penalties but may also be combined with more restrictive measures. These sanctions, applied under administrative law, can impose significant financial burdens on offenders, not only through monetary fines but also via the loss of machinery and production capacity resulting from embargoes and seizures. In some cases, offenders may additionally face civil or criminal charges. Although fines are not the most severe form of punishment, they are the most frequently applied and systematically recorded, serving as a practical proxy for the presence of environmental law enforcement. The ability to impose such penalties depends critically on the timing of enforcement actions. Because Brazilian law allows more binding sanctions when offenders are caught in the act, the use of real-time satellite monitoring through the DETER system has been instrumental in enhancing enforcement effectiveness by enabling prompt response and increasing the likelihood of catching violators red-handed.

Evidence shows that DETER has been effective in curbing deforestation. \citet{assuncao2023deter} document a significant reduction in forest loss inside the Amazon biome following the introduction of real-time satellite monitoring. Beyond environmental outcomes, enforcement efforts may also affect broader dimensions of state presence and local behavior. As noted by \citet{dechezlepretre2017impacts}, environmental policies often generate spillover effects on institutional quality, economic incentives, and conflict dynamics. In this sense, DETER serves not only as a monitoring tool but also as a mechanism through which the state reasserts its authority in contested territories.

\pagebreak

\section{Data Construction} \label{AppData}

\setcounter{table}{0}
\renewcommand\thetable{B.\arabic{table}}

\setcounter{figure}{0}
\renewcommand\thefigure{B.\arabic{figure}}

\setcounter{equation}{0}
\renewcommand\theequation{B.\arabic{equation}}

\setcounter{theorem}{0}
\renewcommand\thetheorem{B.\arabic{theorem}}

\setcounter{proposition}{0}
\renewcommand\theproposition{B.\arabic{proposition}}

\setcounter{corollary}{0}
\renewcommand\thecorollary{B.\arabic{corollary}}

\setcounter{assumption}{0}
\renewcommand\theassumption{B.\arabic{assumption}}

\setcounter{definition}{0}
\renewcommand\thedefinition{B.\arabic{definition}}

\setcounter{Lemma}{0}
\renewcommand\theLemma{B.\arabic{Lemma}}

This appendix provides additional details on the construction, sources, and limitations of the variables used in the empirical analysis. While the main text presents the core concepts and justifications, this section includes more technical discussions and documentation to ensure transparency and replicability.

\subsection{Homicide Rate}
Measuring crime in remote and underserved regions such as the Brazilian Amazon poses significant challenges due to limitations in administrative records and local state capacity. Deaths occurring in isolated areas or among vulnerable populations — especially when not accompanied by a formal death certificate — may be underreported.

Despite these limitations, the  Ministry of Health estimates indicate that SIM captures over 96\% of deaths nationwide, and more then 90\% in the North region \citep{ministeriosaude2011sim}. This high coverage makes SIM reliable for population-level analyses
\citep{cunha2017assessment}. \citet{mathers2005counting} also highlight the quality of SIM data not only for the count of deaths, but also for the consistent use of the International Statistical Classification of Diseases and Related Health Problems, 10\textsuperscript{th} Revision (ICD-10).\footnote{
	The ICD-10 is maintained by the World Health Organization and provides a standardized international framework for coding mortality and morbidity data \citep{WHO2019ICD10}.
}

Table \ref{tab:hsi_shares_icd10} reports annual totals and shares of deaths by ICD-10 cause-of-death categories used in the construction of the homicide rate variable. The classification follows the inclusive approach described in Section~\ref{SecData}, which incorporates assaults (X85--Y09), intentional self-harm (X60--X84), and events of undetermined intent (Y10--Y34) to mitigate underreporting and misclassification of violent deaths in official records. The table shows that assaults — corresponding to homicides in the strict sense — consistently account for approximately 70–75\% of all deaths throughout the sample period. The remaining share is divided between intentional self-harm and undetermined intent. This composition confirms that the constructed homicide measure is predominantly driven by cases explicitly recorded as homicides, while still addressing the measurement challenges inherent to violence data in the Brazilian Amazon.

\begin{table}[ht]
	\centering
	\caption{Annual Totals and Shares of Deaths by ICD-10}
	\label{tab:hsi_shares_icd10}

	\resizebox{\linewidth}{!}{%
		\begin{tabular}{@{\extracolsep{0pt}} cccccccc}
			\toprule
			\midrule
			\textbf{Year} & \textbf{Assaults} & \textbf{Intentional self-harm} & \textbf{Undetermined intent} & \textbf{Total} & \textbf{Share of assaults} & \textbf{Share of intentional self-harm} & \textbf{Share of undetermined intent} \\
			& \textbf{(X85--Y09)} & \textbf{(X60--X84)} & \textbf{(Y10--Y34)} &  & \textbf{(\%)} & \textbf{(\%)} & \textbf{(\%)} \\
			\midrule
			2006 & 48,624 & 8,755 & 9,744 & 67,123 & 72.4 & 13.0 & 14.5 \\
			2007 & 48,308 & 9,148 & 12,283 & 69,739 & 69.3 & 13.1 & 17.6 \\
			2008 & 50,909 & 9,305 & 12,704 & 72,918 & 69.8 & 12.8 & 17.4 \\
			2009 & 52,895 & 9,361 & 11,431 & 73,687 & 71.8 & 12.7 & 15.5 \\
			2010 & 51,539 & 9,714 & 10,140 & 71,393 & 72.2 & 13.6 & 14.2 \\
			2011 & 53,762 & 10,088 & 10,143 & 73,993 & 72.7 & 13.6 & 13.7 \\
			2012 & 56,858 & 10,528 & 9,812 & 77,198 & 73.7 & 13.6 & 12.7 \\
			2013 & 59,513 & 10,477 & 9,533 & 79,523 & 74.8 & 13.2 & 12.0 \\
			2014 & 57,543 & 11,011 & 9,598 & 78,152 & 73.6 & 14.1 & 12.3 \\
			2015 & 60,129 & 11,225 & 10,250 & 81,604 & 73.7 & 13.8 & 12.6 \\
			2016 & 63,547 & 11,890 & 9,950 & 85,387 & 74.4 & 13.9 & 11.7 \\
			\midrule
			\bottomrule
		\end{tabular}
	}

	\vspace{0.5em}

	\resizebox{\linewidth}{!}{%
		\parbox{\linewidth}{\tiny
			\textit{Notes:} Columns 2--4 report annual totals of deaths for each ICD-10 category. Column 5 reports the total number of deaths across these three categories. Columns 6--8 present the share of each ICD-10 category relative to the total for the corresponding year, expressed in percent. Source: SIM-DataSUS
		}
	}
\end{table}

\subsection{Law Enforcement}
As \citet{chalfin2017criminal} argue, deforestation-related fines provide concrete evidence that the state was physically present at the scene of the infraction and took action to hold offenders. What matters for this study is the fact that enforcement agents were present and took formal action, not whether the fine was eventually paid. Therefore, using the number of fines issued regardless of payment status is an appropriate proxy for the presence of environmental monitoring.

\subsection{Cloud Coverage}
The additional information on monthly cloud coverage provided by DETER enables researchers to assess variation in enforcement capacity stemming from weather-related visibility constraints. In cases of partial obstruction, the system records which specific regions within each municipality were visible and which remained obscured. However, if visibility is consistently poor and no clear imagery is available for an entire month, the dataset provides no information for that area. Following INPE's official guidance, we treat the DETER dataset as complete and use its cloud coverage variable to quantify periods of low visibility.

To assess whether cloud coverage reflects persistent geographic characteristics or short-term weather shocks, we compute within-municipality autocorrelations of the DETER cloud coverage variable. Autocorrelations are calculated for one-year and two-year lags over the 2006–2016 sample. Table~\ref{tab:acf_cloud_by_muni} reports summary statistics across all the 521 municipalities.

\begin{table}[ht]
	\centering
	\caption{Autocorrelation of DETER Cloud Coverage within Municipalities}
	\label{tab:acf_cloud_by_muni}
	\begin{tabular}{@{\extracolsep{0pt}} lccccc}
		\toprule
		\midrule
		\multicolumn{6}{c}{\textbf{Panel A: Lag 1 (year $t$ with $t-1$)}} \\
		\midrule
		\textbf{Statistic} & \textbf{Mean} & \textbf{Median} & \textbf{Std. Dev.} & \textbf{P25} & \textbf{P75} \\
		\midrule
		Autocorrelation & 0.095 & 0.101 & 0.294 & -0.093 & 0.302 \\
		\midrule
		\multicolumn{6}{c}{\textbf{Panel B: Lag 2 (year $t$ with $t-2$)}} \\
		\midrule
		\textbf{Statistic} & \textbf{Mean} & \textbf{Median} & \textbf{Std. Dev.} & \textbf{P25} & \textbf{P75} \\
		\midrule
		Autocorrelation & 0.160 & 0.170 & 0.357 & -0.121 & 0.452 \\
		\midrule
		\bottomrule
	\end{tabular}

	\vspace{0.5em}

	\parbox{\linewidth}{\footnotesize
		\textit{Notes:} The table reports cross-sectional summaries of within-municipality autocorrelations of DETER cloud coverage. Panel A shows lag-1 autocorrelation between year $t$ and $t-1$; Panel B shows lag-2 autocorrelation between year $t$ and $t-2$. Statistics are computed over the 521 municipalities in the analysis sample. P25 and P75 denote the 25th and 75th percentiles of the autocorrelation distribution, respectively.
	}

\end{table}

The mean autocorrelation at lag 1 is 0.095 (median 0.101), indicating low persistence year to year. At lag 2, the mean increases to 0.160 (median 0.170). The distribution is wide: the 25th percentile is negative for both lags (–0.093 for lag 1; –0.121 for lag 2), and the 75th percentile is 0.302 and 0.452, respectively. These results suggest that cloud coverage varies substantially over time within municipalities and is not dominated by fixed geographic patterns. This supports the use of cloud coverage as a source of short-term, plausibly exogenous variation in enforcement capacity.

\subsection{Other Controls}
To account for potential confounders and improve the credibility of the identification strategy, we include a rich set of time-varying controls capturing satellite observability, weather patterns, and socioeconomic conditions. These controls are designed to absorb unobserved heterogeneity across municipalities and over time, ensuring that the variation in cloud coverage used as an instrument reflects transitory, exogenous shocks to environmental monitoring capacity rather than structural differences in local governance, climate, or economic activity.

Satellite-based controls include the proportion of municipal area covered by clouds and the share classified as non-observable in PRODES imagery. These indicators help account for persistent monitoring limitations that affect both PRODES and DETER systems. By controlling for these obstructions, we reduce the risk that variation in DETER visibility is confounded by broader weaknesses in monitoring capacity. Such weaknesses may reflect structural constraints in state presence or environmental governance and could influence both enforcement intensity and violence.

Weather variables are added to account for climatic variation that may shape both enforcement capacity and local dynamics. Intense rainfall or extreme heat can limit on-the-ground operations and delay responses to deforestation alerts by reducing accessibility and visibility. At the same time, seasonal and annual weather fluctuations may affect agricultural cycles, land-use pressures, and tensions over natural resources. By including precipitation and average temperature, we aim to capture climate shocks that could simultaneously influence environmental enforcement and violence.

Socioeconomic controls are added to account for heterogeneity in institutional presence and crime incentives. These include a commodity price index (reflecting the value of agricultural output), population density (capturing the spatial opportunity for crime), municipal GDP (as a proxy for local development), and education quality (a proxy for human capital). Each of these variables may influence both violence and the effectiveness of environmental enforcement. Together, they help ensure that the instrument is not spuriously correlated with structural conditions that jointly affect the outcome.

\subsubsection{Commodity Price Index}

\citet{assunccao2015deforestation} show that price series published by the Secretariat of Agriculture and Supply of the State of Paraná (SEAB-PR) are highly correlated with average agricultural prices in Amazon municipalities, making them suitable for the construction of the output price series.

To reflect seasonal variation in agricultural incentives, the index is constructed separately for the first and second semesters of each calendar year. The weighted real price for commodity $c$ in municipality $i$ and semester $st$ is defined as:

\begin{equation}
	PW_{c,i,st} = P_{c,st} \cdot  W_{c,i}
\end{equation}

\noindent
where $PW_{c,i,st}$ is the weighted real price of commodity $c$ in municipality $i$ during semester/year $st$; $P_{c,st}$ is the real price of commodity $c$ during semester/year $st$, and $W_{c,i}$ is the weight for commodity $c$ in  municipality $i$.

After constructing the weighted series for each commodity, we aggregate them into a composite index that summarizes broader market incentives. The commodity index for municipality $i$ in year $t$ is the average of weighted real prices for all commodities across the two semesters of the year.

\subsection{Descriptive Statistics}

\begin{table}[H]
	\centering
	\caption{Main Variables Descriptive Statistics}
	\label{tab:descriptive_statistics}

	\resizebox{\linewidth}{!}{%
		\begin{tabular}{@{\extracolsep{0pt}} l|l|c|cccccccccccc}
			\toprule
			\midrule
			\textbf{Variable} & \textbf{Statistic} & \textbf{Full Sample} & \textbf{2006} & \textbf{2007} & \textbf{2008} & \textbf{2009} & \textbf{2010} & \textbf{2011} & \textbf{2012} & \textbf{2013} & \textbf{2014} & \textbf{2015} & \textbf{2016} \\

			\midrule

			\textbf{Homicide} & \hspace{1em} mean & 16.87 & 11.74 & 12.87 & 14.26 & 15.66 & 16.12 & 16.19 & 17.59 & 18.76 & 19.02 & 21.05 & 22.29 \\
			& \hspace{1em} sd & 73.37 &48.16 & 53.55 & 58.61 & 66.76 & 73.29 & 75.99 & 77.13 & 79.45 & 82.01 & 87.14 & 91.7 \\

			\midrule

			\textbf{Homicide} & \hspace{1em} mean  & 28.16 & 22.95 & 23.3 & 26.12 & 28.08 & 25.46 & 26.18 & 28.92 & 29.64 & 30.84 & 33.31 & 34.94 \\
			\textbf{Rate} & \hspace{1em} sd & 25.64 & 24.42 & 24.08 & 26.88 & 26.5 & 22.52 & 23.91 & 24.83 & 24.02 & 26.1 & 26.76 & 28.75 \\

			\midrule
			\textbf{Environmental} & \hspace{1em} mean  & 9.87 & 12.72 & 11.15 & 16.25 & 11.61 & 9.81 & 10.72 & 6.11 & 8.80 & 6.86 & 10.96 & 3.63 \\
			\textbf{Fines} & \hspace{1em} sd & 28.25 & 26.85 & 23.85 & 37.27 & 32.74 & 23.25 & 26.73 & 16.19 & 30.91 & 24.36 & 41.01 & 13.15 \\

			\midrule

			\textbf{DETER Cloud} & \hspace{1em} mean  & 0.46 & 0.37 & 0.65 & 0.49 & 0.58 & 0.49 & 0.50 & 0.35 & 0.37 & 0.45 & 0.48 & 0.39 \\
			\textbf{Coverage} & \hspace{1em} sd & 0.23 & 0.06 & 0.16 & 0.23 & 0.23 & 0.25 & 0.20 & 0.20 & 0.21 & 0.27 & 0.24 & 0.27 \\

			\midrule

			\textbf{PRODES Cloud} & \hspace{1em} mean  & 578.87 & 68.02 & 376.33 & 568.6 & 441.75 & 434.12 & 827.65 & 557.99 & 585.36 & 1237.18 & 783.31 & 487.27 \\
			\textbf{Coverage} & \hspace{1em} sd & 2572.89 & 262.13 & 1447.33 & 2403.74 & 1804.06 & 1393.36 & 3311.98 & 2879.49 & 2125.07 & 4737.32 & 3023.03 & 1886.78 \\

			\midrule

			\textbf{PRODES} & \hspace{1em} mean  & 19.02 & 47.52 & 46.64 & 47.45 & 21.71 & 9.27 & 7.66 & 7.62 & 7.13 & 7.26 & 6.97 & 0.00 \\
			\textbf{Non-Observable} & \hspace{1em} sd & 263.25 & 261.91 & 262.33 & 231.46 & 37.93 & 36.02 & 35.82 & 34.19 & 33.9 & 34.03 & 0.00 & 156.92 \\

			\midrule

			\textbf{Precipitation} & \hspace{1em} mean  & 7.00 & 6.25 & 7.49 & 7.06 & 7.41 & 7.39 & 6.52 & 7.08 & 6.91 & 7.03 & 7.16 & 6.68 \\
			& \hspace{1em} sd    & 11.05 & 13.49 & 12.33 & 13.20 & 13.54 & 11.70 & 12.46 & 12.47 & 12.67 & 12.49 & 12.45 & 12.55 \\

			\midrule

			\textbf{Temperature} & \hspace{1em} mean  & 26.14 & 26.21 & 26.03 & 26.23 & 25.81 & 26 & 26.52 & 26.21 & 26.12 & 26.2 & 25.96 & 26.211 \\
			& \hspace{1em} sd & 1.26 & 1.17 & 1.22 & 1.13 & 1.28 & 1.21 & 1.32 & 1.21 & 1.28 & 1.3 & 1.38 & 1.26 \\

			\midrule

			\textbf{Commodity} & \hspace{1em} mean  & 6.08 & 4.51 & 4.89 & 5.73 & 6.03 & 5.71 & 6.41 & 6.22 & 5.97 & 6.45 & 7.55 & 7.41 \\
			\textbf{Price Index} & \hspace{1em} sd & 6.89 & 5.02 & 5.42 & 6.34 & 6.72 & 6.32 & 7.15 & 6.93 & 6.55 & 7.11 & 8.52 & 8.31 \\

			\midrule

			\textbf{GDP} & \hspace{1em} mean  & 551.36 & 284.53 & 317.21 & 374.31 & 395.91 & 481.40 & 565.57 & 617.14 & 691.93 & 741.74 & 770.20 & 824.97 \\
			& \hspace{1em} sd & 2896.87 & 1665.22 & 1817.97 & 2037.89 & 2165.75 & 2629.90 & 2979.29 & 3089.17 & 3426.05 & 3614.02 & 3595.65 & 3757.66 \\

			\midrule

			\textbf{Population} & \hspace{1em} mean & 39.48 & 35.45 & 36.65 & 37.17 & 38.26 & 38.80 & 39.29 & 40.65 & 41.21 & 41.75 & 42.28 & 42.78 \\
			& \hspace{1em} sd & 121.44 & 110.32 & 113.38 & 114.96 & 116.82 & 118.39 & 119.86 & 124.73 & 126.50 & 128.23 & 129.92 & 131.57 \\

			\midrule

			\textbf{Population} & \hspace{1em} mean  & 28.52 & 26.12 & 26.97 & 27.39 & 27.67 & 28.06 & 28.43 & 29.11 & 29.47 & 29.82 & 30.16 & 30.48 \\
			\textbf{Density} & \hspace{1em} sd & 155.80 & 148.21 & 151.99 & 154.76 & 149.77 & 151.83 & 153.83 & 157.03 & 158.98 & 160.85 & 162.65 & 164.42 \\

			\midrule

			\textbf{IDEB} & \hspace{1em} mean  & 4.00 & 3.16 & 3.39 & 3.64 & 3.86 & 4.05 & 4.18 & 4.24 & 4.27 & 4.31 & 4.39 & 4.55 \\
			& \hspace{1em} sd & 0.94 & 0.67 & 0.67 & 0.68 & 0.69 & 0.63 & 0.64 & 0.71 & 0.88 & 1.08 & 1.19 & 1.12 \\

			\midrule
			\bottomrule
		\end{tabular}
	}

\end{table}

\textit{Notes of Table \ref{tab:descriptive_statistics}:} The table reports municipality-level means and standard deviations. Variable labels, units, and sources are as follows. Homicide: number of homicide deaths, Mortality Information System (SIM-DataSUS); Homicide Rate: homicide per 100,000 inhabitants, SIM-DataSUS and Brazilian Institute for Geography and Statistics (IBGE); Environmental Fines: number of deforestation-fines, Brazilian Institute for the Environment and Renewable Natural Resources (IBAMA); DETER cloud coverage: ratio of cloud to municipal area, Real-Time System for Detection of Deforestation (DETER) from the Brazilian Institute for Space Research (INPE); PRODES cloud coverage: km\textsuperscript{2}, PRODES/INPE; PRODES non-observable: km\textsuperscript{2}, PRODES/INPE; Precipitation: 10\textsuperscript{3}mm, Matsuura and Willmott (2018a); Temperature: °C, Matsuura and Willmott (2018b); Commodity Index: weighted real price index, IBGE; GDP: BRL1,000,000, IBGE; Population: number of 1,000 inhabitants, IBGE; Population density: ratio of population to municipal area, IBGE; IDEB: composite score, National Institute for Educational Studies and Research Anísio Teixeira (INEP). See Section~\ref{SecData} for details on variable construction.

\pagebreak

\section{Heterogeneity by Death Category}
\label{app_sec:others_ICD}

\setcounter{table}{0}
\renewcommand\thetable{C.\arabic{table}}

\setcounter{figure}{0}
\renewcommand\thefigure{C.\arabic{figure}}

\setcounter{equation}{0}
\renewcommand\theequation{C.\arabic{equation}}

\setcounter{theorem}{0}
\renewcommand\thetheorem{C.\arabic{theorem}}

\setcounter{proposition}{0}
\renewcommand\theproposition{C.\arabic{proposition}}

\setcounter{corollary}{0}
\renewcommand\thecorollary{C.\arabic{corollary}}

\setcounter{assumption}{0}
\renewcommand\theassumption{C.\arabic{assumption}}

\setcounter{definition}{0}
\renewcommand\thedefinition{C.\arabic{definition}}

\setcounter{Lemma}{0}
\renewcommand\theLemma{C.\arabic{Lemma}}

In this appendix, we explore the robustness of our results to alternative and more disaggregated homicide classifications. Specifically, we consider (i) assaults committed with firearms (ICD-10 X93--X95), (ii) assaults committed with sharp or blunt objects (ICD-10 X99--Y00), (iii) deaths resulting from legal intervention (ICD-10 Y35), and (iv) deaths resulting from intentional self-harm (ICD-10 X60--X84). These categories are motivated by the literature and by concerns raised in previous discussions: firearm-related deaths, assaults with physical objects and intentional self-harm capture different types of violence, while legal intervention provides a measure of fatalities caused by state agents in contexts where lethal force is officially sanctioned. Although enforcement in the Amazon is not primarily conducted by conventional police forces but by environmental agencies such as IBAMA, these distinctions allow us to test whether changes in enforcement affect specific causes of violent mortality.

Table~\ref{tab:homicide_other_specification} reports the 2SLS estimates for these alternative outcomes. The results indicate no statistically significant association between environmental enforcement and firearm-related homicides (Column (1)). By contrast, the coefficient for assaults with sharp or blunt objects (Column (2)) is negative and statistically significant at the 5\% level: one additional fine is associated with a reduction of 0.33 homicides per 100,000 inhabitants, corresponding to a 3.7\% decrease relative to the average homicide rate in this category. This effect is sizeable, implying that moving from the 25th to the 75th percentile in fines (roughly nine additional fines) is associated with a reduction of about one-third of the average homicide rate due to assaults with sharp or blunt objects. Additionally, the coefficient for deaths due to legal intervention (Column (3)) is small and statistically insignificant, suggesting that increased environmental enforcement is not associated with higher rates of fatalities in confrontations with state agents. This result is consistent with the institutional context in which IBAMA, rather than police forces, carries out most enforcement activities, and confrontations rarely escalate into lethal violence. Lastly, the coefficient for deaths due to intentional self-harm (Column (4)) is small and statistically insignificant, suggesting that increased environmental enforcement is not associated with higher rates of suicides.

\begin{table}[htp]
	\centering
	\caption{2SLS --- Estimates under Alternative Homicide Classifications}
	\label{tab:homicide_other_specification}
	\begin{tabularx}{\textwidth}{lCCCC}
		\toprule
		\toprule

		& \multicolumn{4}{c}{Homicide Rate} \\
		& (1) & (2) & (3) & (4) \\

		\midrule

		Lagged Enforcement  & -0.132 & -0.330$^{**}$ & 0.002 & -0.048 \\
		& (0.194) & (0.165)       & (0.008) & (0.078)\\

		& & & & \\

		\midrule

		Average homicide rate across municipalities  & 11.65 & 8.89 & 0.04 & 1.95 \\

		& & & & \\

		\midrule

		FE: Municipality \& Year & \checkmark & \checkmark & \checkmark & \checkmark \\

		& & & & \\

		All controls & \checkmark & \checkmark & \checkmark & \checkmark \\

		& & & & \\

		Observations & 5,210 & 5,210 & 5,210 & 5,210 \\
		Municipalities & 521 & 521 & 521 & 521 \\

		\bottomrule

	\end{tabularx}

	\vspace{0.1em}

	\resizebox{\linewidth}{!}{%
		\parbox{\linewidth}{\footnotesize
			\textit{Notes:} 2SLS coefficients are estimated based on Equation~\eqref{eq:second_stage} from Section~\ref{SecEmpirical}. Column (1) focuses on assault by firearm (ICD-10 X93--X95), Column (2) focuses on assault by sharp or blunt object (ICD-10 X99--Y00), Column (3) focuses on legal intervention (ICD-10 Y35), and Column (4) focuses on intentional self-harm (ICD-10 X60--X84). ``Homicide Rate'' is the number of homicides per 100,000 inhabitants. ``Lagged Enforcement'' refers to the total number of fines issued and serves as a proxy for law enforcement effectiveness. The set of control variables contains PRODES cloud coverage and non-observable (satellite); precipitation and temperature (weather); and commodity index, GDP, population density and Ideb scores (socioeconomic). The dataset is a municipality-by-year panel covering the period 2006--2016. The sample includes all Amazon biome municipalities that exhibited variation in forest cover during the sample period and for which deforestation data are available. Standard errors are clustered at the municipality level and reported in parentheses. Significance: $^{*}p<0.1$, $^{**}p<0.05$, $^{***}p<0.01$.
		}
	}

\end{table}

\pagebreak

\section{Heterogeneity by Sex, Race and Age Group}
\label{app_sec:sex_race}

\setcounter{table}{0}
\renewcommand\thetable{D.\arabic{table}}

\setcounter{figure}{0}
\renewcommand\thefigure{D.\arabic{figure}}

\setcounter{equation}{0}
\renewcommand\theequation{D.\arabic{equation}}

\setcounter{theorem}{0}
\renewcommand\thetheorem{D.\arabic{theorem}}

\setcounter{proposition}{0}
\renewcommand\theproposition{D.\arabic{proposition}}

\setcounter{corollary}{0}
\renewcommand\thecorollary{D.\arabic{corollary}}

\setcounter{assumption}{0}
\renewcommand\theassumption{D.\arabic{assumption}}

\setcounter{definition}{0}
\renewcommand\thedefinition{D.\arabic{definition}}

\setcounter{Lemma}{0}
\renewcommand\theLemma{D.\arabic{Lemma}}

A natural question is whether the estimated effects of environmental enforcement on homicides may vary across demographic groups. For example, \citet{bfps2025} show national evidence that lethal violence is disproportionately concentrated among young men and African Brazilians.

In line with IBGE (Brazilian Census Bureau) classification, we group individuals into three broad racial categories: White and Yellow, African Brazilian (Black and Pardo), and Indigenous. The African Brazilian group combines individuals classified as ``Preto'' (Black) and Pardo. While ``Preto'' refers to individuals self-identified as Black, Pardo is a census category used in Brazil to capture mixed-race individuals with diverse backgrounds, reflecting the country’s history of racial mixture \citep{ibge2013caracteristicas}. We adopt the term African Brazilian to designate this combined group throughout the analysis.

In this appendix, we analyze differences by sex (Men v. Women), race (White and Yellow v. African Brazilian v. Indigenous) and age groups (0--15 years v. 15--25 years v. 25--35 years v. 36+ years). To this end, we divide this appendix into three sections: (i) sex, (ii) race and (iii) age groups. In each section, we first present descriptive statistics on the distribution of homicides and then estimate the same regression specification as in our main analysis.

\subsection{Sex}
\label{app_subsec:sex}

Table~\ref{tab:hsi_shares_sex} reports the annual distribution of homicide victims by sex between 2006 and 2016. The descriptive statistics reveal that the vast majority of homicide victims are men, consistently representing around 87--88\% of all cases throughout the period, while women account for roughly 12\% and cases of unknown sex remain negligible. This stability over time is in line with the broader literature documenting that lethal violence in Brazil is overwhelmingly concentrated among men \citep{bfps2025}.

\begin{table}[ht]
	\centering
	\caption{Annual Totals and Shares of Deaths by Sex}
	\label{tab:hsi_shares_sex}

	\resizebox{\linewidth}{!}{%
		\begin{tabular}{@{\extracolsep{0pt}} cccccccccc}
			\toprule
			\midrule
			{Year} & {Men} & {Women} & {Unknown} & {Total} & {Share of} & {Share of} & {Share of} \\
			& & & & & { men (\%)} & {women (\%)} & {unknown (\%)} \\
			{(1)} & {(2)} & {(3)} & {(4)} & {(5)} & {(6)} & {(7)} & {(8)} \\
			\midrule
			2006 & 59,068 & 7,965 & 90 & 67,123 & 88.0 & 11.9 & 0.1\\
			2007 & 61,235 & 8,412 & 92 & 69,739 & 87.8 & 12.1 & 0.1\\
			2008 & 63,926 & 8,900 & 92 & 72,918 & 87.7 & 12.2 & 0.1\\
			2009 & 64,505 & 9,087 & 95 & 73,687 & 87.5 & 12.3 & 0.1\\
			2010 & 62,160 & 9,155 & 78 & 71,393 & 87.1 & 12.8 & 0.1\\
			2011 & 64,456 & 9,436 & 101 & 73,993 & 87.1 & 12.8 & 0.1\\
			2012 & 67,605 & 9,474 & 119 & 77,198 & 87.6 & 12.3 & 0.2\\
			2013 & 69,977 & 9,404 & 142 & 79,523 & 88.0 & 11.8 & 0.2\\
			2014 & 68,546 & 9,481 & 125 & 78,152 & 87.7 & 12.1 & 0.2\\
			2015 & 71,921 & 9,546 & 137 & 81,604 & 88.1 & 11.7 & 0.2\\
			2016 & 75,546 & 9,690 & 151 & 85,387 & 88.5 & 11.3 & 0.2\\
			\midrule
			\bottomrule
		\end{tabular}
	}

	\vspace{0.5em}

	\resizebox{\linewidth}{!}{%
		\parbox{\linewidth}{\footnotesize
			\textit{Notes:} Columns (2)--(4) report annual totals of deaths for each sex category. Column (5) reports the total number of deaths across these three categories. Columns (6)--(8) present the share of each sex category relative to the total for the corresponding year, expressed in percent. Source: SIM-DataSUS
		}
	}
\end{table}

Table~\ref{tab:homicide_sex} presents the 2SLS estimates by sex. The results indicate that environmental enforcement significantly reduces homicides among men: the coefficient is –0.75 and statistically significant at the 5\% level. Given an average male homicide rate of 25.05 per 100,000 inhabitants, this estimate implies that the issuance of one additional fine is associated with a reduction of about 3.0\% relative to the mean male homicide rate. By contrast, the effect for women is close to zero (0.005) and not statistically significant, consistent with the expectation that environmental enforcement primarily affects types of violence linked to illegal economic activities, which disproportionately involve men \citep{soares2021ilegalidade, bfps2025}.

\begin{table}[htp]
	\centering
	\caption{2SLS --- Estimates Conditioning on Sex}
	\label{tab:homicide_sex}
	\begin{tabularx}{\textwidth}{lCC}
		\toprule
		\toprule

		& \multicolumn{2}{c}{Homicide Rate} \\
		& Men & Women \\
		& (1) & (2) \\

		\midrule

		Lagged Enforcement  & -0.754$^{**}$   & 0.005 \\
		& (0.343)          & (0.091) \\

		& &  \\

		\midrule

		Average homicide rate across municipalities  & 25.05 & 3.02 \\

		& &  \\

		\midrule

		FE: Municipality \& Year & \checkmark & \checkmark  \\

		& &\\

		All controls & \checkmark & \checkmark \\

		& & \\

		Observations & 5,210 & 5,210 \\
		Municipalities & 521 & 521  \\

		\bottomrule

	\end{tabularx}

	\vspace{0.1em}

	\resizebox{\linewidth}{!}{%
		\parbox{\linewidth}{\footnotesize
			\textit{Notes:} 2SLS coefficients are estimated based on Equation~\eqref{eq:second_stage} from Section~\ref{SecEmpirical}. Column (1) reports results for men and Column (2) for women. ``Homicide Rate'' is the number of homicide victims of the corresponding sex per 100,000 inhabitants. ``Lagged Enforcement'' refers to the total number of fines issued and serves as a proxy for law enforcement effectiveness. The set of control variables contains PRODES cloud coverage and non-observable (satellite); precipitation and temperature (weather); and commodity index, GDP, population density and Ideb scores (socioeconomic). The dataset is a municipality-by-year panel covering the period 2006-2016. The sample includes all Amazon biome municipalities that exhibited variation in forest cover during the sample period and for which deforestation data are available. Standard errors are clustered at the municipality level and reported in parentheses. Significance: $^{*}p<0.1$, $^{**}p<0.05$, $^{***}p<0.01$.
		}
	}

\end{table}

\subsection{Race}
\label{app_subsec:race}

Table~\ref{tab:hsi_shares_race_aggregated} reports the annual distribution of homicide victims by aggregated race groups. The data show that the majority of homicide victims are consistently classified as African Brazilian, representing around 58--69\% of all cases over the period, while White and Yellow account for roughly 30--35\% and Indigenous individuals for less than 1\%. This concentration among the African Brazilian population mirrors the persistent racial disparities in homicide victimization documented by \citet{bfps2025} at the national level.

\begin{table}[htp]
	\centering
	\caption{Annual Totals and Shares of Deaths by Aggregated Race Groups}
	\label{tab:hsi_shares_race_aggregated}

	\resizebox{\linewidth}{!}{%
		\begin{tabular}{@{\extracolsep{0pt}} c|ccccc|cccc}
			\toprule
			\midrule
			& \multicolumn{5}{c|}{Levels} & \multicolumn{4}{c}{Shares}\\
			{Year} & {White and} & {African} & {Indigenous} & {Unknown} & {Total} & {White and} & {African} & {Indigenous} & {Unknown} \\
			& Yellow & Brazilian &  & & & {Yellow (\%)} & {Brazilian (\%)} & {(\%)} & {(\%)} \\
			(1) & (2) & (3) & (4) & (5) & (6) & (7) & (8) & (9) & (10)\\
			\midrule
			2006 & 23,840 & 38,745 & 229 & 4,309 & 67,123 & 35.5 & 57.7 & 0.3 & 6.4\\
			2007 & 23,923 & 41,231 & 266 & 4,319 & 69,739 & 34.3 & 59.1 & 0.4 & 6.2\\
			2008 & 24,757 & 43,797 & 270 & 4,094 & 72,918 & 34.0 & 60.1 & 0.4 & 5.6\\
			2009 & 24,109 & 44,915 & 221 & 4,442 & 73,687 & 32.7 & 61.0 & 0.3 & 6.0\\
			2010 & 23,460 & 43,560 & 244 & 4,129 & 71,393 & 32.9 & 61.0 & 0.3 & 5.8\\
			2011 & 23,074 & 46,275 & 299 & 4,345 & 73,993 & 31.2 & 62.5 & 0.4 & 5.9\\
			2012 & 23,704 & 48,442 & 331 & 4,721 & 77,198 & 30.7 & 62.8 & 0.4 & 6.1\\
			2013 & 23,776 & 50,737 & 343 & 4,667 & 79,523 & 29.9 & 63.8 & 0.4 & 5.9\\
			2014 & 23,244 & 50,466 & 379 & 4,063 & 78,152 & 29.7 & 64.6 & 0.5 & 5.2\\
			2015 & 23,854 & 53,418 & 347 & 3,985 & 81,604 & 29.2 & 65.5 & 0.4 & 4.9\\
			2016 & 24,169 & 58,727 & 418 & 2,073 & 85,387 & 28.3 & 68.8 & 0.5 & 2.4\\
			\midrule
			\bottomrule
		\end{tabular}
	}

	\vspace{0.5em}

	\resizebox{\linewidth}{!}{%
		\parbox{\linewidth}{\footnotesize
			\textit{Notes:} Columns (2)--(5) report annual totals of deaths for aggregated race groups. Column (6) reports the total number of deaths. Columns (7)--(10) present the share of each aggregated group relative to the total for the corresponding year, expressed in percent. Source: SIM-DataSUS
		}
	}
\end{table}

Table~\ref{tab:homicide_race_groups} presents the 2SLS estimates by race group. The results indicate that environmental enforcement significantly reduces homicides among the African Brazilian population: the coefficient equals –0.742 and is statistically significant at the 5\% level. Given an average homicide rate of 21.84 per 100,000 inhabitants in this group, the effect implies that the issuance of one additional fine is associated with a reduction of about 3.4\% relative to the mean homicide rate. By contrast, the effect is small and statistically insignificant for the White and Yellow population (–0.075) and for Indigenous individuals (0.061). These findings underscore that the deterrent effects of enforcement are concentrated among African Brazilian individuals, the group disproportionately exposed to violence in the Amazon region.

\begin{table}[ht]
	\centering
	\caption{2SLS --- Estimates Conditioning on Race}
	\label{tab:homicide_race_groups}
	\begin{tabularx}{\textwidth}{lCCC}
		\toprule
		\toprule

		& \multicolumn{3}{c}{Homicide Rate} \\
		& White and Yellow & African Brazilian & Indigenous \\

		\midrule

		Lagged Enforcement  & -0.075 & -0.742$^{**}$  & 0.061   \\
		& (0.127) & (0.309) & (0.060) \\

		& & & \\

		\midrule

		Average homicide rate across municipalities  & 4.49 & 21.84 & 1.11   \\

		& & & \\

		\midrule

		FE: Municipality \& Year & \checkmark & \checkmark & \checkmark \\

		& & & \\

		All controls & \checkmark & \checkmark & \checkmark \\

		& & & \\

		Observations & 5,210 & 5,210 & 5,210 \\
		Municipalities & 521 & 521 & 521 \\

		\bottomrule

	\end{tabularx}

	\vspace{0.1em}

	\resizebox{\linewidth}{!}{%
		\parbox{\linewidth}{\footnotesize
			\textit{Notes:} 2SLS coefficients are estimated based on Equation~\eqref{eq:second_stage} from Section~\ref{SecEmpirical}. Column (1) reports results for the white and yellow population, Column (2) for the black and mixed-race population, and Column (3) for the indigenous population. ''Homicide Rate'' is the number of homicides victims in each municipality per 100,000 inhabitants. ''Lagged Enforcement'' refers to the total number of fines issued and serves as a proxy for law enforcement effectiveness. The set of control variables contains PRODES cloud coverage and non-observable (satellite); precipitation and temperature (weather); and commodity index, GDP, population density and Ideb scores (socioeconomic). The dataset is a municipality-by-year panel covering the period 2006-2016. The sample includes all Amazon biome municipalities that exhibited variation in forest cover during the sample period and for which deforestation data are available. Standard errors are clustered at the municipality level and reported in parentheses. Significance: $^{*}p<0.1$, $^{**}p<0.05$, $^{***}p<0.01$.
		}
	}

\end{table}

\subsection{Age Groups}
\label{app_subsec:age}

Table~\ref{tab:hsi_shares_age} reports the annual distribution of homicide victims by age group between 2006 and 2016. The descriptive statistics reveal that the majority of homicide victims are 26 years or older, consistently representing around 59--63\% of all cases throughout the period.

\begin{table}[ht]
	\centering
	\caption{Annual Totals and Shares of Deaths by Age Group}
	\label{tab:hsi_shares_age}

	\resizebox{\linewidth}{!}{%
		\begin{tabular}{@{\extracolsep{0pt}} cccccccccccc}
			\toprule
			\midrule
			{Year} & {0-15} & {16-25} & {26-35} & {36+} & {Unknown} & {Total} & {Share of} & {Share of} & {Share of} & {Share of} & {Share of}\\
			&  &  &  &  &  &  & {0-15 (\%)} & {16-25 (\%)} & {26-35 (\%)} & {36+ (\%)} & {Unknown (\%)}\\
			{(1)} & {(2)} & {(3)} & {(4)} & {(5)} & {(6)} & {(7)} & {(8)} & {(9)} & {(10)} & {(11)} & {(12)} \\
			\midrule
			2006 & 2,226 & 22,636 & 17,321 & 23,412 & 1,528 & 67,123 & 3.3 & 33.7 & 25.8 & 34.9 & 2.3\\
			2007 & 2,398 & 23,409 & 17,776 & 24,468 & 1,688 & 69,739 & 3.4 & 33.6 & 25.5 & 35.1 & 2.4\\
			2008 & 2,388 & 23,949 & 18,991 & 25,813 & 1,777 & 72,918 & 3.3 & 32.8 & 26.0 & 35.4 & 2.4\\
			2009 & 2,419 & 23,865 & 19,643 & 26,122 & 1,638 & 73,687 & 3.3 & 32.4 & 26.7 & 35.4 & 2.2\\
			2010 & 2,321 & 22,824 & 18,681 & 26,085 & 1,482 & 71,393 & 3.3 & 32.0 & 26.2 & 36.5 & 2.1\\
			2011 & 2,471 & 23,844 & 19,480 & 26,685 & 1,513 & 73,993 & 3.3 & 32.2 & 26.3 & 36.1 & 2.0\\
			2012 & 2,515 & 25,485 & 19,949 & 27,700 & 1,549 & 77,198 & 3.3 & 33.0 & 25.8 & 35.9 & 2.0\\
			2013 & 2,680 & 26,602 & 20,702 & 28,041 & 1,498 & 79,523 & 3.4 & 33.5 & 26.0 & 35.3 & 1.9\\
			2014 & 2,477 & 25,258 & 20,120 & 28,982 & 1,315 & 78,152 & 3.2 & 32.3 & 25.7 & 37.1 & 1.7\\
			2015 & 2,515 & 26,878 & 20,207 & 30,779 & 1,225 & 81,604 & 3.1 & 32.9 & 24.8 & 37.7 & 1.5\\
			2016 & 2,468 & 28,767 & 21,264 & 31,624 & 1,264 & 85,387 & 2.9 & 33.7 & 24.9 & 37.0 & 1.5\\
			\midrule
			\bottomrule
		\end{tabular}
	}

	\vspace{0.5em}

	\resizebox{\linewidth}{!}{%
		\parbox{\linewidth}{\tiny
			\textit{Notes:} Columns (2)--(6) report annual totals of deaths for each age group. Column (7) reports the total number of deaths across these four categories. Columns (8)- (12) present the share of each age group relative to the total for the corresponding year, expressed as a percentage. Source: SIM-DataSUS
		}
	}
\end{table}

Table~\ref{tab:homicide_age_new} presents the 2SLS estimates by age group. The results indicate that environmental enforcement significantly reduces homicides only among individuals whose ages are between 26 and 35 years: the coefficient is –0.35 and statistically significant at the 5\% level. Given an average homicide rate of 7.59 per 100,000 inhabitants for this age group, this estimate implies that the issuance of one additional fine is associated with a reduction of about 4.6\% relative to the mean homicide rate for this age group. By contrast, the effect for younger individuals is close to zero (0.05 for the 0-15 years group and -0.08 for the 16--25 years group) and not statistically significant.

\begin{table}[htp]
	\centering
	\caption{2SLS --- Estimates Conditioning on Age Group}
	\label{tab:homicide_age_new}
	\begin{tabularx}{\textwidth}{lCCCC}
		\toprule
		\toprule

		& \multicolumn{4}{c}{Homicide Rate} \\
		& 0--15 & 16--25 & 26-35 & 36+ \\

		\midrule

		Lagged Enforcement  & 0.050  & -0.083  & -0.349$^{**}$ & -0.329    \\
		& (0.076) & (0.143) & (0.162)       & (0.202) \\

		& & & &  \\

		\midrule

		Average homicide rate across municipalities  & 1.22 & 8.06 & 7.59 & 10.56   \\

		& & & &   \\

		\midrule

		FE: Municipality \& Year & \checkmark & \checkmark & \checkmark & \checkmark  \\

		& & & &  \\

		All controls & \checkmark & \checkmark & \checkmark & \checkmark \\

		& & & &  \\

		Observations & 5,210 & 5,210 & 5,210 & 5,210 \\
		Municipalities & 521 & 521 & 521 & 521  \\

		\bottomrule

	\end{tabularx}

	\vspace{0.1em}

	\resizebox{\linewidth}{!}{%
		\parbox{\linewidth}{\footnotesize
			\textit{Notes:} 2SLS coefficients are estimated based on Equation~\eqref{eq:second_stage} from Section~\ref{SecEmpirical}. Column (1) reports results for the 0--15 age group, Column (2) for ages 16--25, Column (3) for ages 26--35, and Column (4) for ages 36 and above. ``Homicide Rate'' is the number of homicide victims of the corresponding sex per 100,000 inhabitants. ``Lagged Enforcement'' refers to the total number of fines issued and serves as a proxy for law enforcement effectiveness. The set of control variables contains PRODES cloud coverage and non-observable (satellite); precipitation and temperature (weather); and commodity index, GDP, population density and Ideb scores (socioeconomic). The dataset is a municipality-by-year panel covering the period 2006-2016. The sample includes all Amazon biome municipalities that exhibited variation in forest cover during the sample period and for which deforestation data are available. Standard errors are clustered at the municipality level and reported in parentheses. Significance: $^{*}p<0.1$, $^{**}p<0.05$, $^{***}p<0.01$.
		}
	}

\end{table}

\pagebreak

\section{Controlling for Conservation Policies}
\label{sec:robustness_conservation_policies}

\setcounter{table}{0}
\renewcommand\thetable{E.\arabic{table}}

\setcounter{figure}{0}
\renewcommand\thefigure{E.\arabic{figure}}

\setcounter{equation}{0}
\renewcommand\theequation{E.\arabic{equation}}

\setcounter{theorem}{0}
\renewcommand\thetheorem{E.\arabic{theorem}}

\setcounter{proposition}{0}
\renewcommand\theproposition{E.\arabic{proposition}}

\setcounter{corollary}{0}
\renewcommand\thecorollary{E.\arabic{corollary}}

\setcounter{assumption}{0}
\renewcommand\theassumption{E.\arabic{assumption}}

\setcounter{definition}{0}
\renewcommand\thedefinition{E.\arabic{definition}}

\setcounter{Lemma}{0}
\renewcommand\theLemma{E.\arabic{Lemma}}

In addition to the robustness checks in the main text, we explore an alternative specification that incorporates policy controls for major federal conservation programs, such as the establishment of protected areas and environmental pacts. Although potentially relevant, these controls are excluded from the baseline specification due to concerns about reverse causality and policy endogeneity. For instance, enforcement and conservation policies may be jointly determined in response to local violence or illegal deforestation trends. This issue would bias our results because we do not have an instrument for these additional conservation policies.

Conservation policies may influence where enforcement occurs and how violence evolves across space. These policies can affect land-use conflict, increase state presence, or alter incentives for illegal activity in ways that overlap with formal enforcement. Although not part of the main specification, we include controls for conservation policies to ensure that the estimated effects of law enforcement are not confounded by overlapping conservation initiatives.

We include two key conservation policy variables in the analysis. First, we use the share of municipal territory designated as a sustainable-use protected area, specifically Environmental Protection Areas (Áreas de Proteção Ambiental, APAs), based on spatial overlap with units listed in Brazil’s National Registry of Conservation Units (CNUC). APAs are legal conservation units that allow for sustainable land use but restrict deforestation and land conversion. Second, we include an indicator for whether the municipality was designated as a Priority Municipality under the federal government’s conservation strategy launched in 2008. This policy targeted municipalities with the highest deforestation rates with intensified enforcement actions, including embargoes, fines, and credit restrictions.\footnote{
	As shown by \citet{assunccao2019getting}, this prioritization was effective in significantly reducing deforestation.
}  These conservation policy controls help ensure that other concurrent conservation initiatives do not confound the estimated effect of environmental enforcement on violence.

Table~\ref{tab:conservation_policies} reports the estimated second-stage coefficients when we control for these additional conservation policies. Column (1) replicates the main specification while Column (2) includes the two conservation policy controls. The estimated effect of environmental enforcement on homicide remains negative and statistically significant, although the effect becomes slightly more negative (–0.748) and significance at the 10\% level only. The first-stage F-statistic remains above conventional thresholds (14.25), mitigating concerns about weak instruments. These results reinforce the conclusion that the observed relationship between law enforcement and violence is not simply driven by the spatial targeting of conservation policy.

\begin{table}[htb]
	\centering
	\caption{2SLS --- Conservation Policies Control}
	\label{tab:conservation_policies}
	\begin{tabularx}{\textwidth}{lCC}
		\toprule
		\midrule

		& \multicolumn{2}{c}{Homicide Rate} \\
		& (1) & (2) \\

		\midrule

		Lagged Enforcement & -0.728$^{**}$  & -0.748$^{*}$  \\
		& (0.351)       & (0.385) \\

		\midrule

		First Stage F-statistic & 16.40 & 14.25 \\

		FE: Municipality \& Year & \checkmark & \checkmark \\

		All controls & \checkmark & \checkmark \\
		Conservation Policies & \xmark & \checkmark \\

		Observations & 5,210 & 5,210 \\
		Municipalities & 521 & 521 \\

		\bottomrule

	\end{tabularx}

	\vspace{0.1em}

	\resizebox{\linewidth}{!}{%
		\parbox{\linewidth}{\footnotesize
			\textit{Notes:} 2SLS coefficients are estimated based on Equation~\eqref{eq:second_stage} from Section~\ref{SecEmpirical}. Column (1) is the main specification; Column (2) includes two conservation policies as additional controls. ``Homicide Rate'' refers to the number of homicides per 100,000 inhabitants. ``Lagged Enforcement'' refers to the total number of fines issued and serves as a proxy for law enforcement. The set of control variables contains PRODES cloud coverage and non-observable; precipitation and temperature; and commodity index, GDP, population density and Ideb scores. The dataset is a municipality-by-year panel covering the period 2006-2016. The sample includes all Amazon biome municipalities that exhibited variation in forest cover during the sample period and for which deforestation data are available. Standard errors are clustered at the municipality level. Significance: $^{*}p<0.1$, $^{**}p<0.05$, $^{***}p<0.01$.
		}
	}

\end{table}

\end{document}